\documentclass{article}%
\usepackage{amsmath}
\usepackage{amsfonts}
\usepackage{amssymb}
\usepackage{graphicx}%
\providecommand{\U}[1]{\protect\rule{.1in}{.1in}}
\newtheorem{theorem}{Theorem}
\newtheorem{acknowledgement}[theorem]{Acknowledgement}

\newtheorem{claim}[theorem]{Claim}

\newtheorem{remark}[theorem]{Remark}

\begin{document}

\title{Elko Spinor Fields and Massive Magnetic Like Monopoles}
\author{E. Capelas de Oliveira, W. A. Rodrigues Jr. and J. Vaz Jr.\\Institute of Mathematics, Statistics and Scientific Computation\\IMECC-UNICAMP\\13083-859 Campinas SP, Brazil\\walrod@ime.unicamp.br, capelas@ime.unicamp.br, vaz@ime.unicamp.br}
\date{November 16 2013}
\maketitle

\begin{abstract}
In this paper we recall that by construction Elko spinor fields of
$\boldsymbol{\lambda}$ and $\boldsymbol{\rho}$ types satisfy a coupled system
of first order partial differential equations (\emph{c}\textit{sfopde}) that
once interacted leads to Klein-Gordon equations for the $\boldsymbol{\lambda}$
and $\boldsymbol{\rho}$ type fields. Since the \emph{c}\textit{sfopde} is the
basic one and since the Klein-Gordon equations for $\boldsymbol{\lambda}$ and
$\boldsymbol{\rho}$ possess solutions that are \textit{not} solutions of the
\emph{c}\textit{sfopde} for $\boldsymbol{\lambda}$ and $\boldsymbol{\rho}$ we
infer that it is legitimate to attribute to those fields mass dimension $3/2$
(as\ is the case of Dirac spinor fields) and \emph{not} mass dimension $1$ as
previously suggested in recent literature (see list of references).\ A proof
of this fact is offered by deriving the \emph{c}\textit{sfopde for the
}$\boldsymbol{\lambda}$ and $\boldsymbol{\rho}$ from a Lagrangian where these
fields\ have indeed mass dimension $3/2$. Taking seriously the view that Elko
spinor fields due to its special properties given by their bilinear invariants
may be the description of some kind of particles in the real world a question
then arises: what is the physical meaning of these fields? Here we proposed
that the fields $\boldsymbol{\lambda}$ and $\boldsymbol{\rho}$ serve the
purpose of building the fields $\mathcal{K}\in\mathcal{C}\ell^{0}%
\mathbf{(}M\mathbf{,}\eta\mathbf{)}\otimes\mathbb{R}_{1,3}^{0}$ and
$\mathcal{M}\in\sec\mathcal{C}\ell^{0}\mathbf{(}M\mathbf{,}\eta\mathbf{)}%
\otimes\mathbb{R}_{1,3}^{0}$ (see Eq.(\ref{ceA0})). These fields are
\emph{electrically neutral }but carry \emph{magnetic} like charges\ which
permit them to couple to a $su(2)\simeq spin_{3,0}\subset\mathbb{R}_{3,0}^{0}$
valued potential $\mathcal{A}\in\sec%
{\textstyle\bigwedge\nolimits^{1}}
T^{\ast}M\otimes\mathbb{R}_{3,0}^{0}$. If the field $\mathcal{A}$ is of short
range the particles described by the $\mathcal{K}$\textbf{ }and $\mathcal{M}$
fields may be interacting and forming condensates of zero spin particles
analogous to dark matter, in the sense that they do not couple with the
electromagnetic field (generated by charged particles) and are thus invisible.
Also, since according to our view the Elko spinor fields as well as the
$\mathcal{K}$\textbf{ }and $\mathcal{M}$ fields are of mass dimension $3/2$ we
show how to calculate the correct propagators for the $\mathcal{K}$\textbf{
}and $\mathcal{M}$ fields. We discuss also the main difference between Elko
and Majorana spinor fields, which are kindred since both belong to class five
in Lounesto classification of spinor fields. Most of our presentation uses the
representation of spinor fields in the Clifford bundle formalism, which makes
very clear the meaning of all calculations.

\end{abstract}

\section{Introduction}

Elko spinor fields have been introduced in \cite{ag2005,ag2005a} as \emph{dual
helicity} eigenspinors of the charge conjugation operator satisfying
Klein-Gordon equation and carrying according to the authors of
\cite{ag2005,ag2005a} mass dimension $1$ instead of mass dimension $3/2$
carried by Dirac spinor fields. A considerable number of \ interesting papers
have been published in the literature on these extraordinary objects in the
past few years. In particular, according to the theory in
\cite{ag2005,ag2005a} the anticommutator of an elko spinor field with its
conjugate momentum is nonlocal and it is claimed that the theory possess an
axis of locality which implies also that the theory of elko spinor fields
break Lorentz invariance. We shall discuss this issue in the Appendix B which
according our view is an odd feature of the theory in \cite{ag2005,ag2005a}.
We recall in section 2 that differently from the theory in
\cite{ag2005,ag2005a} where a second quantized elko spinor field satisfies a
Klein-Gordon equation (instead of a Dirac equation) the\ classical elko spinor
fields of $\boldsymbol{\lambda}$ and $\boldsymbol{\rho}$ types satisfy by
their construction a \emph{c}\textit{sfopde }that is Lorentz invariant. It is
thus possible to construct (what we believe) is a more reasonable theory where
those fields, as we will show play a key role. The \emph{c}\textit{sfopde}
once interacted leads to Klein-Gordon equations for the $\boldsymbol{\lambda}$
and $\boldsymbol{\rho}$ type fields. However, since the \emph{c}%
\textit{sfopde} is the basic one and since the Klein-Gordon equations for
$\boldsymbol{\lambda}$ and $\boldsymbol{\rho}$ possess solutions that are
\textit{not} solutions of the \emph{c}\textit{sfopde} for $\lambda$ and $\rho$
we think that it is not necessary to get the field equations for
$\boldsymbol{\lambda}$ and $\boldsymbol{\rho}$ from a Lagrangian where those
fields have mass dimension $1$ as in \cite{ag2005,ag2005a}. Indeed, we claim
that we can attribute mass dimension of $3/2$ for these fields as is the case
of Dirac spinor fields. A proof of this fact is offered by deriving in Section
3 the \emph{c}\textit{sfopde for }$\boldsymbol{\lambda}$ and $\boldsymbol{\rho
}$ from a Lagrangian where these fields\ have mass dimension $3/2$. This, fact
is to be contrasted with the quantum theory of these fields as presented in
\cite{ag2005,ag2005a,als2010,als2011,rh2007,rh2010,siro2009} (and references
therein), namely that elko fields have mass dimension 1.\ 

Taking seriously the view that elko spinor fields due to the special
properties given by their bilinear invariants may be the description of some
kind of particles in the real world a question then arises: what is the
physical meaning of these fields?

In what follows we propose that the fields $\lambda$ and $\rho$\ (the
representatives in the Clifford bundle $\mathcal{C}\ell\mathbf{(}%
M\mathbf{,}\eta\mathbf{)}$ of the covariant spinor fields $\boldsymbol{\lambda
}$ and $\boldsymbol{\rho}$ ) serve the purpose of building Clifford valued
multiform fields, i.e., $\mathcal{K}\in\mathcal{C}\ell^{0}\mathbf{(}%
M\mathbf{,}\eta\mathbf{)}\otimes\mathbb{R}_{1,3}^{0}$ and $\mathcal{M}\in
\sec\mathcal{C}\ell^{0}\mathbf{(}M\mathbf{,}\eta\mathbf{)}\otimes
\mathbb{R}_{1,3}^{0}$ (see Eq.(\ref{ceA1})). These fields are
\emph{electrically neutral }but carry \emph{magnetic} like charges\ which
permit that they couple to a $su(2)\simeq spin_{3,0}\subset\mathbb{R}%
_{1,3}^{0}$ valued potential $\mathcal{A}\in\sec%
{\textstyle\bigwedge\nolimits^{1}}
T^{\ast}M\otimes spin_{3,0}$. If the field $\mathcal{A}$ is of short range the
particles described by the $\mathcal{K}$\textbf{ }and $\mathcal{M}$ may be
interacting and forming a system of spin zero particles with zero magnetic
like charge and eventually form condensates something analogous to dark
matter, in the sense that they do not couple with the electromagnetic field
and are thus invisible.

We observe that elko and Majorana fields are in class 5 of Lounesto
classification \cite{lounesto} and although an elko spinor field does
\emph{not} satisfy the Dirac equation as correctly claimed in
\cite{ag2005,ag2005a}, a Majorana spinor field $\boldsymbol{\psi}_{\mathbf{M}%
}:M\rightarrow\mathbb{C}^{4}$ \ which is a dual helicity object according to
some authors (see e.g., \cite{maggiore}) \emph{does} satisfy the Dirac
equation. However this statement is not correct. However an operator
(quantum)Majorana field $\boldsymbol{\psi}_{\mathbf{M}}$ can satisfy Dirac
equation if it is not a dual helicity object ( see Section 5.3). Also, even at
a \textquotedblleft classical level\textquotedblright\ a Majorana spinor field
satisfies Dirac equation\ if for any\ $x\in M$ \ their components take values
in a Grassmann algebra. Also, differently from the case of elko spinor fields
some authors claim that Majorana fields are \textit{not} dual helicities
objects \cite{ag2005}, a statement that is correct only for Majorana quantum
fields as constructed in Section 5.3. For a Majorana field \ (even at
\textquotedblleft classical level\textquotedblright) whose components take
values in a Grassmann algebra the statement is not correct.

Finally, since according to our findings the elko spinor fields as well as the
fields $\mathcal{K}$ and $\mathcal{M}$ are of mass dimension $3/2$ we show in
Section 6 how to calculate the correct propagators for $\mathcal{K}$\textbf{
}and $\mathcal{M}$. We also show that the causal propagator for the covariant
$\boldsymbol{\lambda}$ and $\boldsymbol{\rho}$ \ fields is simply the standard
Feynman propagator of Dirac theory.

In presenting the above results we use the representation of spinor fields in
the Clifford bundle formalism (CBF) \cite{mr2004,r2004,rodcap2007}. This is
briefly recalled in section 2 where a useful translation for the standard
matrix formalism.\footnote{If more details are need the reader may find the
necessary help in \cite{rodcap2007}.} to the CBF is given. The CBF makes all
calculations easy and transparent and in particular permits to infer
\cite{roro2006} in a while that elko spinor fields are class 5 spinor fields
in Lounesto classification \cite{lounesto, roro2006}.

\section{Description of Spinor Fields in the Clifford Bundle}

Let $(M\simeq\mathbb{R}^{4},\mathbf{\eta},D,\tau_{\mathbf{\eta}})$ be the
Minkowski spacetime structure where $\mathbf{\eta}\in\sec T_{0}^{2}M$ is
Minkowski metric and $D$ is the Levi-Civita connection of $\mathbf{\eta}$.
Also, $\tau_{\mathbf{\eta}}\in\sec%
{\textstyle\bigwedge\nolimits^{4}}
T^{\ast}M$ defines an orientation. We denote by $\eta\in\sec T_{2}^{0}M$ the
metric of the cotangent bundle. It is defined as follows. Let $\langle x^{\mu
}\rangle$ be coordinates for \ $M$ in the Einstein-Lorentz-Poincar\'{e} gauge.
Let $\langle\boldsymbol{e}_{\mu}=\partial/\partial x^{\mu}\rangle$ a basis for
$TM$ and $\langle\gamma^{\mu}=dx^{\mu}\rangle$ the corresponding dual basis
for $T^{\ast}M$, i.e., $\gamma^{\mu}(\boldsymbol{e}_{\alpha})=\delta_{\alpha
}^{\mu}$. Then, if $\mathbf{\eta}=\eta_{\mu\nu}\gamma^{\mu}\otimes\gamma^{\nu
}$ then $\eta=\eta^{\mu\nu}\boldsymbol{e}_{\mu}\otimes\boldsymbol{e}_{\nu}$,
where the matrix with entries $\eta_{\mu\nu}$ and the one with entries
$\eta^{\mu\nu}$ are the equal to the diagonal matrix $\mathrm{diag}%
(1,-1,-1,-1)$. If $a,b\in\sec%
{\textstyle\bigwedge\nolimits^{1}}
T^{\ast}M$ \ we write $a\cdot b=\eta(a,b)$. We also denote by $\langle
\gamma_{\mu}\rangle$ the reciprocal basis of $\langle\gamma^{\mu}=dx^{\mu
}\rangle$, which satisfies $\gamma^{\mu}\cdot\gamma_{\nu}=\delta_{\nu}^{\mu}$.

We denote the Clifford bundle of differential forms\footnote{We recall that
$\mathcal{C\ell}(T_{x}^{\ast}M,\eta)\simeq\mathbb{R}_{1,3}$ the so-called
spacetime algebra. Also the even subalgebra of $\mathbb{R}_{1,3}$ denoted
$\mathbb{R}_{1,3}^{0}$ is isomorphic to te Pauli algebra $\mathbb{R}_{3,0}$,
i.e., $\mathbb{R}_{1,3}^{0}\simeq\mathbb{R}_{3,0}$. The even subalgebra of the
Pauli algebra $\mathbb{R}_{3,0}^{0}:=\mathbb{R}_{3,0}^{00}$ is the quaternion
algebra $\mathbb{R}_{0,2}$, i.e., $\mathbb{R}_{0,2}\simeq\mathbb{R}_{3,0}^{0}%
$. Moreover we have the identifications: $Spin_{1,3}^{0}\simeq Sl(2,\mathbb{C}%
)$, $Spin_{3,0}\simeq SU(2)$. For the Lie algebras of these groups we have
$spin_{1,3}^{0}\simeq sl(2,\mathbb{C})$,$\ su(2)\simeq spin_{3,0}$. The
important fact to keep in mind for the understanding of some of the
identificastions we done below is that $Spin_{1,3}^{0},spin_{1,3}^{0}%
\subset\mathbb{R}_{3,0}\subset\mathbb{R}_{1,3}$ and $Spin_{3,0},spin_{3,0}%
\subset\mathbb{R}_{0,2}\subset\mathbb{R}_{1,3}^{0}\subset\mathbb{R}_{1,3}$. If
more details are need the read should consult, e.g., \cite{rodcap2007}.} by
$\mathcal{C\ell}(M,\eta)$ and use notations and conventions in what follows as
in \cite{rodcap2007} and recall the fundamental relation
\begin{equation}
\gamma^{\mu}\gamma^{\nu}+\gamma^{\nu}\gamma^{\mu}=2\eta^{\mu\nu}. \label{1}%
\end{equation}

As well known all (covariant) spinor fields carrying a $(1/2,0)\oplus(0,1/2)$
representation of \textrm{Spin}$_{1,3}^{0}\simeq\mathrm{Sl}(2,\mathbb{C})$
belongs to one of the six Lounesto classes \cite{lounesto}. As well known a
$(1/2,0)\oplus(0,1/2)$ spinor field in Minkowski spacetime is an equivalence
class of triplets $(\boldsymbol{\psi},\Sigma,\Xi)$ where for each $x\in M,$
$\boldsymbol{\psi}(x)\in\mathbb{C}^{4}$, $\Sigma$ is an orthonormal coframe
and $\Xi=u\in\mathrm{Spin}_{1,3}^{0}(M,\eta)\subset\mathcal{C\ell}(M,\eta)$ is
a spinorial frame. If we fix a fiducial global coframe $\Sigma_{0}%
=\langle\mathring{\gamma}^{\mu}\rangle$ and take, e.g., $\Xi_{0}=u_{0}%
=1\in\mathrm{Spin}_{1,3}^{0}(M,\eta)\subset\mathcal{C\ell}(M,\eta)$ the
triplet $(\boldsymbol{\psi}_{0},\Sigma_{0},\Xi_{0})$ is equivalent to
$(\psi,\Sigma,\Xi)$ \ if $\gamma^{\mu}=\Lambda_{\nu}^{\mu}\mathring{\gamma
}^{\nu}=(\pm u)\gamma^{\mu}(\pm u^{-1})$ and $\boldsymbol{\psi}%
(x)=S(u)\boldsymbol{\psi}_{0}(x)$ where $S(u)$ is the standard $(1/2,0)\oplus
(0,1/2)$ matrix representation of $\mathrm{Sl}(2,\mathbb{C})$. Dirac gamma
matrices in standard and Weyl representations will be denoted by
$\boldsymbol{\gamma}^{\mu}$ and $\boldsymbol{\gamma}^{\mu^{\prime}}$ and are
not to be confused with the $\gamma^{\mu}\in\sec%
{\textstyle\bigwedge\nolimits^{1}}
T^{\ast}M$ $\hookrightarrow\mathcal{C\ell}(M,\eta)$. As well known the gamma
matrices satisfy $\boldsymbol{\gamma}^{\mu}\boldsymbol{\gamma}^{\nu
}+\boldsymbol{\gamma}^{\nu}\boldsymbol{\gamma}^{\mu}=2\eta^{\mu\nu}$ and
$\boldsymbol{\gamma}^{\mu^{\prime}}\boldsymbol{\gamma}^{\prime\nu
}+\boldsymbol{\gamma}^{\prime\nu}\boldsymbol{\gamma}^{\prime\mu}=2\eta^{\mu
\nu}$. The relation between the $\boldsymbol{\gamma}^{\mu}\boldsymbol{\ }$and
the $\boldsymbol{\gamma}^{\prime\mu}$ is given by
\begin{equation}
\boldsymbol{\gamma}^{\prime\mu}=S\boldsymbol{\gamma}^{\mu}S^{-1}
\label{gamma0}%
\end{equation}
where\footnote{We will supress the writing of the $4\times4$ and the
$2\times2$ unity matrices when no confusion arises.}
\begin{equation}
S=\frac{1}{\sqrt{2}}\left(
\begin{array}
[c]{cc}%
\mathbf{1} & \mathbf{1}\\
\mathbf{1} & -\mathbf{1}%
\end{array}
\right)  . \label{tmatrix}%
\end{equation}

A representation of a $(1/2,0)\oplus(0,1/2)$ spinor field in the Clifford
bundle is an equivalence class of triplets $(\psi,\Sigma,\Xi)$ where $\psi
\in\sec\mathcal{C\ell}^{0}(M,\eta)$ (the even subbundle of $\sec
\mathcal{C\ell}(M,\eta)$), $\Sigma$ is an orthonormal coframe and $\Xi
_{u}=u\in\mathrm{Spin}_{1,3}^{0}(M,\eta)\subset\mathcal{C\ell}(M,\eta)$ is a
spinorial frame. If \ we fix a fiducial global coframe $\Sigma_{0}%
=\langle\Gamma^{\mu}\rangle$ and take $\Xi_{u_{0}}=u_{0}=1\in\sec
\mathrm{Spin}_{1,3}^{0}(M,\eta)\subset\sec\mathcal{C\ell}(M,\eta)$ the triplet
$(\psi_{0},\Sigma_{0},\Xi_{0})$ is equivalent to\footnote{Take notice that
$(\psi,\Sigma,\Xi_{u})$ is not equivalent to $(\psi,\Sigma,\Xi_{-u})$ even if
$(u)\boldsymbol{\gamma}^{\mu}(u^{-1})=(-u)\boldsymbol{\gamma}^{\mu}(-u^{-1}%
)$.} $(\psi,\Sigma,\Xi_{u})$ if $\gamma^{\mu}=\Lambda_{\nu}^{\mu}\Gamma^{\nu
}=(u)\boldsymbol{\gamma}^{\mu}(u^{-1})$ and $\psi=\psi_{0}u^{-1}$. Field
$\psi$ is called an operator spinor field and the operator spinor fields
belonging to Lounesto classes $1,2,3$ are also known as Dirac-Hestenes spinor fields.

If $\boldsymbol{\gamma}^{\mu}$, $\mu=0,1,2,3$ are the Dirac gamma matrices in
the \emph{standard representation} and $\langle\gamma_{\mu}\rangle$ are as
introduced above, we define%
\begin{align}
\sigma_{k}  &  :=\gamma_{k}\gamma_{0}\in\sec%
{\textstyle\bigwedge\nolimits^{2}}
T^{\ast}M\hookrightarrow\sec\mathcal{C\ell}^{0}(M,\eta)\text{, }%
k=1,2,3,\label{2}\\
\mathbf{i}  &  =\gamma_{5}:=\gamma_{0}\gamma_{1}\gamma_{2}\gamma_{3}\in\sec%
{\textstyle\bigwedge\nolimits^{4}}
T^{\ast}M\hookrightarrow\sec\mathcal{C\ell}(M,\eta),\\
\boldsymbol{\gamma}_{5}  &  :=\boldsymbol{\gamma}_{0}\boldsymbol{\gamma}%
_{1}\boldsymbol{\gamma}_{2}\boldsymbol{\gamma}_{3}\in\mathrm{Mat}%
(4,\mathbb{C).}%
\end{align}

Then, to the covariant spinor $\boldsymbol{\psi}:M\rightarrow\mathbb{C}^{4}$
(in standard representation of the gamma matrices) where ($i=\sqrt{-1}$,
$\boldsymbol{\phi},\boldsymbol{\varsigma}:M\rightarrow\mathbb{C}^{2}$)%
\begin{equation}
\boldsymbol{\psi}=\left(
\begin{array}
[c]{c}%
\boldsymbol{\phi}\\
\boldsymbol{\varsigma}%
\end{array}
\right)  =\left(
\begin{array}
[c]{c}%
\left(
\begin{array}
[c]{c}%
m^{0}+im^{3}\\
-m^{2}+im^{1}%
\end{array}
\right) \\
\left(
\begin{array}
[c]{c}%
n^{0}+in^{3}\\
-n^{2}+in^{1}%
\end{array}
\right)
\end{array}
\right)  , \label{3}%
\end{equation}
there corresponds the operator spinor field $\psi\in\sec\mathcal{C\ell}%
^{0}(M,\eta)$ given by%
\begin{equation}
\psi=\phi+\varsigma\sigma_{3}=(m^{0}+m^{k}\mathbf{i}\sigma_{k})+(n^{0}%
+n^{k}\mathbf{i}\sigma_{k})\sigma_{3}. \label{4}%
\end{equation}
We then have the useful formulas in Eq.(\ref{5}) below that one can use to
immediately translate results of the standard matrix formalism in the language
of the Clifford bundle formalism and vice-versa\footnote{$\tilde{\psi}$ is the
reverse of $\psi$. If $A_{r}\in\sec%
{\textstyle\bigwedge\nolimits^{r}}
T^{\ast}M\hookrightarrow\sec\mathcal{C\ell}(M,\eta)$ then $\tilde{A}%
_{r}=(-1)^{\frac{r}{2}(r-1)}A_{r}$.}
\begin{align}
\boldsymbol{\gamma}_{\mu}\boldsymbol{\psi}  &  \leftrightarrow\gamma_{\mu}%
\psi\gamma_{0},\nonumber\\
i\boldsymbol{\psi}  &  \leftrightarrow\psi\gamma_{21}=\psi\mathbf{i}\sigma
_{3},\nonumber\\
i\boldsymbol{\gamma}_{5}\boldsymbol{\psi}  &  \leftrightarrow\psi\sigma
_{3}=\psi\gamma_{3}\gamma_{0},\nonumber\\
\boldsymbol{\bar{\psi}}  &  =\boldsymbol{\psi}^{\dagger}\boldsymbol{\gamma
}^{0}\leftrightarrow\tilde{\psi},\nonumber\\
\boldsymbol{\psi}^{\dagger}  &  \leftrightarrow\gamma_{0}\tilde{\psi}%
\gamma_{0},\nonumber\\
\boldsymbol{\psi}^{\ast}  &  \leftrightarrow-\gamma_{2}\psi\gamma_{2}.
\label{5}%
\end{align}

\begin{remark}
Note that $\boldsymbol{\gamma}_{\mu},i\mathbf{1}_{4}$ and the operations
$\overline{}$ and $\dagger$ are for each $x\in M$ mappings $\mathbb{C}%
^{4}\rightarrow\mathbb{C}^{4}$. Then they are represented in the Clifford
bundles formalism by extensor fields \emph{\cite{rodcap2007}} which maps
$\mathcal{C\ell}^{0}(M,\eta)$ $\rightarrow\mathcal{C\ell}^{0}(M,\eta)$. Thus,
to the operator $\boldsymbol{\gamma}_{\mu}$ there corresponds an extensor
field, call it \underline{$\boldsymbol{\gamma}$}$_{\mu}:\mathcal{C\ell}%
^{0}(M,\eta)$ $\rightarrow\mathcal{C\ell}^{0}(M,\eta)$ such that
\underline{$\boldsymbol{\gamma}$}$_{\mu}\psi=\gamma_{\mu}\psi\gamma_{0}$.
\end{remark}

Using the above dictionary the standard Dirac equation\footnote{$\partial
_{\mu}:=\frac{\partial}{\partial x^{\mu}}$.} for a Dirac spinor field
$\boldsymbol{\psi}:M\rightarrow\mathbb{C}^{4}$
\begin{equation}
i\boldsymbol{\gamma}^{\mu}\partial_{\mu}\boldsymbol{\psi}-m\boldsymbol{\psi}=0
\label{6}%
\end{equation}
translates immediately in the so-called Dirac-Hestenes equation, i.e.,
\begin{equation}
\boldsymbol{\partial}\psi\gamma_{21}-m\psi\gamma_{0}=0. \label{7}%
\end{equation}

\begin{remark}
In \emph{Eq.(\ref{7}) the operator} $\boldsymbol{\partial}$ acts on
$\mathcal{C\in}\sec\mathcal{C\ell}(M,\eta)$ \emph{(}when using the basis
introduced above\emph{)} as\footnote{The symbols $\lrcorner$ and $\wedge$
denote respctivley the leftcontraction and the exterior products in
$\mathcal{C\ell}(M,\eta)$.}%
\begin{equation}
\boldsymbol{\partial}\mathcal{C}:\mathcal{=\gamma}^{\mu}\lrcorner
(\partial_{\mu}\mathcal{C)+\gamma}^{\mu}\wedge(\partial_{\mu}\mathcal{C)}
\label{7mm}%
\end{equation}

\end{remark}

\begin{remark}
\label{multi}It is sometimes useful, in particular when studying solutions for
the Dirac-Hestenes equation to consider the Clifford bundle of multivector
fields $\mathcal{C\ell}(M,\boldsymbol{\eta})$. We will write $\check{\psi}%
\in\sec\mathcal{C\ell}(M,\boldsymbol{\eta})$ for the sections of the
$\mathcal{C\ell}(M,\boldsymbol{\eta})$ bundle. The Dirac-Hestenes equation in
$\mathcal{C\ell}(M,\boldsymbol{\eta})$ is.%
\begin{equation}
\boldsymbol{\check{\partial}}\check{\psi}\boldsymbol{e}_{21}-m\check{\psi
}\boldsymbol{e}_{0}=0. \label{7m}%
\end{equation}
where $\boldsymbol{e}_{\mu}\boldsymbol{e}_{\nu}+\boldsymbol{e}_{\nu
}\boldsymbol{e}_{\mu}=2\eta_{\mu\nu}$ and $\boldsymbol{\check{\partial}%
:=e}^{\mu}\partial_{\mu}$ with $\boldsymbol{e}^{\mu}:=\eta^{\mu\nu}$
and\emph{(}when using the basis introduced above\emph{)}%
\begin{equation}
\boldsymbol{\check{\partial}}\mathcal{\check{C}}:=\boldsymbol{e}^{\mu
}\lrcorner(\partial_{\mu}\mathcal{\check{C})+}\boldsymbol{e}^{\mu}%
\wedge(\partial_{\mu}\mathcal{\check{C})}, \label{7mmm}%
\end{equation}
for $\mathcal{\check{C}}\in\sec\mathcal{C\ell}(M,\boldsymbol{\eta})$. Keep in
mind that in definition of $\boldsymbol{\check{\partial}}$ the $\boldsymbol{e}%
^{\mu}$ are not supposed to act as a derivatives operators, i.e.,
$\boldsymbol{e}^{\mu}\lrcorner(\partial_{\mu}\mathcal{\check{C})}$
\emph{(}respectively $\boldsymbol{e}^{\mu}\wedge(\partial_{\mu}\mathcal{\check
{C})}$) is the left contraction$\ $of $\boldsymbol{e}^{\mu}$ with
$\partial_{\mu}\mathcal{\check{C}}$ \emph{(}respectively, the exterior product
of $\boldsymbol{e}^{\mu}$ with $\partial_{\mu}\mathcal{\check{C}}$\emph{)}.
\end{remark}

The basic positive and negative energy solutions of Eq.(\ref{6}) which are
eigenspinors of the helicity operator are \cite{sch}
\begin{equation}
\mathbf{u}^{(1)}(\mathbf{p)}e^{-ip_{\mu}x^{\mu}},\text{ \ \ }\mathbf{u}%
^{(2)}(\mathbf{p)}e^{-ip_{\mu}x^{\mu}},\text{ \ \ }\mathbf{v}^{(1)}%
(\mathbf{p)}e^{ip_{\mu}x^{\mu}},\text{ \ \ }\mathbf{v}^{(2)}(\mathbf{p)}%
e^{ip_{\mu}x^{\mu}}. \label{7aa}%
\end{equation}
The $\mathbf{u}^{(\alpha)}(\mathbf{p)}$ and $\mathbf{v}^{(\alpha)}%
(\mathbf{p)}$ ($\alpha=1,2$) are eigenspinors of the parity
operator\footnote{The parity operator acting on covariant spinor fields is
defined as in \cite{ag2005}, i.e., $\mathbf{P}=i\gamma^{0}\mathcal{R}$,
\ where $\mathcal{R}$ changes $\mathbf{p\mapsto-p}$ and changes the
eingenvalues of the helicity operator. For other possibilities for the parity
operator, see e.g., page 50 of \cite{blp}.} $\mathbf{P}$, i.e.,
\begin{equation}
\mathbf{Pu}^{(\alpha)}(\mathbf{p)}=\mathbf{u}^{(\alpha)}(\mathbf{p),}\text{
\ \ }\mathbf{Pv}^{(\alpha)}(\mathbf{p)}=\mathbf{v}^{(\alpha)}(\mathbf{p),}
\label{parity}%
\end{equation}
which makes Dirac equation invariant under a parity
transformation\footnote{For an easy and transparent way to see this rresult
see Appendix .}. These fields are represented in the Clifford bundle formalism
by the following operator spinor fields,%
\begin{equation}
u^{(r)}(\mathbf{p)=}L(\mathbf{p)}\varkappa^{(r)},\text{ \ \ }v^{(r)}%
(\mathbf{p)=}L(\mathbf{p)}\varkappa^{(r)}\sigma_{3}, \label{7aaa}%
\end{equation}
where $\varkappa^{(r)}=\{1,-\mathbf{i}\sigma_{2}\}$ and $L(\mathbf{p)}$ is the
following boost operator ($L(\mathbf{p)}\tilde{L}(\mathbf{p)=1}$%
)\footnote{Recall that $p\gamma^{0}=p_{\mu}\gamma^{\mu}\gamma^{0}%
=E+\mathbf{p}$.}
\begin{equation}
L(\mathbf{p)=}\frac{p\gamma^{0}+m}{\sqrt{2m(E+m)}}. \label{7a4}%
\end{equation}

\begin{remark}
Recall that Dirac-Hestenes spinor fields couple to the electromagnetic
potential $A\in\sec%
{\textstyle\bigwedge\nolimits^{1}}
T^{\ast}M\hookrightarrow\sec\mathcal{C\ell}(M,\eta)$ as%
\begin{equation}
\boldsymbol{\partial}\psi\gamma_{21}-m\psi\gamma_{0}+eA\psi=0. \label{7a}%
\end{equation}
As it is well known this equation is invariant under a parity transformation
of the fields $A$ and $\psi$.
\end{remark}

In \cite{ag2005} the following (covariant) self and anti-self dual elko spinor
fields $\boldsymbol{\lambda}_{\{+-\}}^{\prime s,a},$\newline%
$\boldsymbol{\lambda}_{\{-+\}}^{\prime s,a}$ $\boldsymbol{\rho}_{\{+-\}}%
^{\prime s,a},\boldsymbol{\rho}_{\{-+\}}^{\prime s,a}:M\rightarrow
\mathbb{C}^{4}$ which are eigenspinors of the charge conjugation operator
($\mathbf{C}$)\footnote{The conjugation operator used in \cite{ag2005} is
$\mathbf{C}\mathbb{\boldsymbol{\psi}=-\gamma}^{2}\mathbb{\boldsymbol{\psi}%
}^{\ast}$. Using the dictionary given by Eq.(\ref{5}) we find that in the
Clifford bundle formalism we have $\mathbf{C}\psi\mathbb{=-}\psi
\mathbb{\gamma}_{20}$.} are defined \ using the Weyl,(chiral) representation
of the gamma matrices by%

\begin{align}
\boldsymbol{\lambda}_{\{\mp\text{ }\pm\}}^{\prime s}(\mathbf{p)}  &  =\left(
\begin{array}
[c]{c}%
\mathbf{\sigma}_{2}[\mathbf{\phi}_{L}^{\pm}(\mathbf{p})]^{\ast}\\
\mathbf{\phi}_{L}\mathbf{\pm}(\mathbf{p})
\end{array}
\right)  ,\text{ \ \ }\boldsymbol{\lambda}_{\{\mp\text{ }\pm\}}^{\prime
a}(\mathbf{p)=}\left(
\begin{array}
[c]{c}%
-\mathbf{\sigma}_{2}[\mathbf{\phi}_{L}^{\pm}(\mathbf{p})]^{\ast}\\
\mathbf{\phi}_{L}\mathbf{\pm}(\mathbf{p})
\end{array}
\right)  ,\label{deflambda}\\
\boldsymbol{\rho}_{\{\pm\text{ }\mp\}}^{\prime s}(\mathbf{p)}  &  =\left(
\begin{array}
[c]{c}%
\mathbf{\phi}_{R}\mathbf{\pm}(\mathbf{p})\\
-\mathbf{\sigma}_{2}[\mathbf{\phi}_{R}^{\pm}(\mathbf{p})]^{\ast}%
\end{array}
\right)  ,\text{ \ \ }\boldsymbol{\rho}_{\{\pm\text{ }\mp\}}^{\prime
a}(\mathbf{p)=}\left(
\begin{array}
[c]{c}%
\mathbf{\phi}_{R}\mathbf{\pm}(\mathbf{p})\\
\mathbf{\sigma}_{2}[\mathbf{\phi}_{R}^{\pm}(\mathbf{p})]^{\ast}%
\end{array}
\right)  ,
\end{align}
where the $\mathbf{C}\boldsymbol{\lambda}^{\prime s}=+\boldsymbol{\lambda
}^{\prime s}$, $\mathbf{C}\boldsymbol{\lambda}^{\prime a}=-\boldsymbol{\lambda
}^{\prime a}$ and the indices $\{+-\},\{-+\}$ refers to the helicities of the
upper and down components of the elko spinor fields, and where as in
\cite{ag2005} we introduce the following helicity eigenstates\footnote{The
indices $L$ and $R$ in $\mathbf{\phi}_{L}^{\pm}(\mathbf{p)}$ and
$\mathbf{\phi}_{L}^{\pm}(\mathbf{p})$ refer to the fact that these spinors
fields transforms according to the basic non equivalent two dimensional
representation of $Sl(2,\mathbb{C})$.}, $\mathbf{\phi}_{L}^{+}(\mathbf{0})$
and $\mathbf{\phi}_{L}^{-}(\mathbf{0})$ and $\mathbf{\phi}_{R}^{+}%
(\mathbf{0})$ and $\mathbf{\phi}_{R}^{-}(\mathbf{0})$ such that with
$\mathbf{\hat{p}\frac{\mathbf{p}}{\left\vert \mathbf{p}\right\vert }}$ we have%
\begin{align}
\boldsymbol{\sigma\cdot}\frac{\mathbf{p}}{\left\vert \mathbf{p}\right\vert
}\mathbf{\phi}_{L}^{\pm}(\mathbf{0})  &  :=\pm\mathbf{\phi}_{L}^{\pm
}(\mathbf{0}),\text{ \ \ \ }\boldsymbol{\sigma\cdot}\frac{\mathbf{p}%
}{\left\vert \mathbf{p}\right\vert }[\boldsymbol{\sigma}_{2}(\mathbf{\phi}%
_{L}^{\pm}(\mathbf{0}))^{\ast}]=\mp\lbrack\boldsymbol{\sigma}_{2}%
(\mathbf{\phi}_{L}^{\pm}(\mathbf{0}))^{\ast}],\nonumber\\
\boldsymbol{\sigma\cdot}\frac{\mathbf{p}}{\left\vert \mathbf{p}\right\vert
}\mathbf{\phi}_{R}^{\pm}(\mathbf{0})  &  :=\pm\mathbf{\phi}_{R}^{\pm
}(\mathbf{0}),\text{ \ \ \ }\boldsymbol{\sigma\cdot}\frac{\mathbf{p}%
}{\left\vert \mathbf{p}\right\vert }[-\boldsymbol{\sigma}_{2}(\mathbf{\phi
}_{R}^{\pm}(\mathbf{0}))^{\ast}]=\mp\lbrack-\boldsymbol{\sigma}_{2}%
(\mathbf{\phi}_{R}^{\pm}(\mathbf{0}))^{\ast}]. \label{helicity}%
\end{align}
Also recall that being a \emph{general} boost operator in the $D^{1/2,0}%
\oplus$ $D^{0,1/2}$ representation of $Sl(2,\mathbb{C)}$
\begin{equation}
\boldsymbol{K=K}^{1/2,0}\oplus\boldsymbol{K}^{0,1/2}=e^{\frac
{\boldsymbol{\sigma}}{2}\boldsymbol{\cdot}\mathbf{\alpha}}\oplus
e^{-\frac{\boldsymbol{\sigma}}{2}\boldsymbol{\cdot}\mathbf{\alpha}}
\label{boost}%
\end{equation}
we have, e.g., taking $\mathbf{\alpha=p}$
\begin{equation}
\boldsymbol{\lambda}_{\{-+\}}^{\prime s}(\mathbf{p)=}\sqrt{\frac{E+m}{m}%
}\left(  1-\frac{\left\vert \mathbf{p}\right\vert }{E+m}\right)
\boldsymbol{\lambda}_{\{-+\}}^{\prime s}(\mathbf{0),} \label{example}%
\end{equation}
More details, if necessary, may be found in \cite{ag2005}.

\begin{remark}
By dual helicity field we simply mean here that the formulas in
\emph{Eq.(\ref{helicity})} are satisfied. Note that the helicity operator
\emph{(}in both Weyl and standard representation of the gamma matrices\emph{)}
is
\begin{equation}
\mathbf{\Sigma}\boldsymbol{\cdot}\frac{\mathbf{p}}{\left\vert \mathbf{p}%
\right\vert }=\left(
\begin{array}
[c]{cc}%
\boldsymbol{\sigma\cdot}\frac{\mathbf{p}}{\left\vert \mathbf{p}\right\vert } &
\mathbf{0}\\
\mathbf{0} & \boldsymbol{\sigma\cdot}\frac{\mathbf{p}}{\left\vert
\mathbf{p}\right\vert }%
\end{array}
\right)  . \label{heli}%
\end{equation}
$\mathbb{C}^{4}$-valued spinor fields depends for its definition of a choice
of an inertial frame where the momentum of the particle is $(p_{0}%
,\mathbf{p})$. The operator $\emph{(}\boldsymbol{K}^{1/2,0}\oplus
\boldsymbol{K}^{0,1/2})$ commutes with $\boldsymbol{\sigma\cdot}%
\frac{\mathbf{p}}{\left\vert \mathbf{p}\right\vert }$ only if
$\boldsymbol{\sigma}\cdot\boldsymbol{\alpha}$ is proportional to
$\boldsymbol{\sigma\cdot}\frac{\mathbf{p}}{\left\vert \mathbf{p}\right\vert }%
$. So, the statement in \emph{\cite{ag2005}} that the helicity operator
commutes with the boost operator must be qualified. However, it remains true
that $\mathbf{\sigma}_{2}[\mathbf{\phi}_{l}^{+}(\mathbf{p})]^{\ast}$ and
$\mathbf{\phi}_{l}^{+}(\mathbf{p})$ have opposite helicities\emph{ }for
any\emph{ }$\mathbf{p}$.
\end{remark}

\begin{remark}
Recall that, e.g., the $\mathbb{C}^{4}$-valued spinor field
$\boldsymbol{\lambda}_{\{-\text{ }+\}}^{\prime s}(\mathbf{p})$ given in the
Weyl representation of the gamma matrices is represented by
$\boldsymbol{\lambda}_{\{-\text{ }+\}}^{s}(\mathbf{p})$ in the standard
representation of the gamma matrices. We have
\begin{align}
\boldsymbol{\lambda}_{\{-\text{ }+\}}^{s}(\mathbf{p})  &
=S\boldsymbol{\lambda}_{\{-\text{ }+\}}^{\prime s}(\mathbf{p})=\frac{1}%
{\sqrt{2}}\left(
\begin{array}
[c]{cc}%
\boldsymbol{1} & \mathbf{1}\\
\mathbf{1} & \boldsymbol{-1}%
\end{array}
\right)  \left(
\begin{array}
[c]{c}%
\mathbf{\sigma}_{2}[\mathbf{\phi}_{L}^{+}(\mathbf{p})]^{\ast}\\
\mathbf{\phi}_{L}^{\mathbf{+}}(\mathbf{p})
\end{array}
\right) \nonumber\\
&  =\frac{1}{\sqrt{2}}\left(
\begin{array}
[c]{c}%
\mathbf{\sigma}_{2}[\mathbf{\phi}_{L}^{+}(\mathbf{p})]^{\ast}+\mathbf{\phi
}_{L}^{\mathbf{+}}(\mathbf{p})\\
\mathbf{\sigma}_{2}[\mathbf{\phi}_{L}^{+}(\mathbf{p})]^{\ast}-\mathbf{\phi
}_{L}^{\mathbf{+}}(\mathbf{p})
\end{array}
\right)
\end{align}
and then%
\begin{equation}
\mathbf{\Sigma}\cdot\frac{\mathbf{p}}{\left\vert \mathbf{p}\right\vert }%
\frac{1}{\sqrt{2}}\left(
\begin{array}
[c]{c}%
\mathbf{\sigma}_{2}[\mathbf{\phi}_{L}^{+}(\mathbf{p})]^{\ast}+\mathbf{\phi
}_{L}^{\mathbf{+}}(\mathbf{p})\\
\mathbf{\sigma}_{2}[\mathbf{\phi}_{L}^{+}(\mathbf{p})]^{\ast}-\mathbf{\phi
}_{L}^{\mathbf{+}}(\mathbf{p})
\end{array}
\right)  =\frac{1}{\sqrt{2}}\left(
\begin{array}
[c]{c}%
-\mathbf{\sigma}_{2}[\mathbf{\phi}_{L}^{+}(\mathbf{p})]^{\ast}+\mathbf{\phi
}_{L}^{\mathbf{+}}(\mathbf{p})\\
-\mathbf{\sigma}_{2}[\mathbf{\phi}_{L}^{+}(\mathbf{p})]^{\ast}-\mathbf{\phi
}_{L}^{\mathbf{+}}(\mathbf{p})
\end{array}
\right)  .
\end{equation}

\end{remark}

\begin{remark}
Recall that, e.g., the $\mathbb{C}^{4}$-valued spinor field
$\boldsymbol{\lambda}_{\{-\text{ }+\}}^{\prime s}(\mathbf{p})$ given in the
Weyl representation of the gamma matrices is represented by
$\boldsymbol{\lambda}_{\{-\text{ }+\}}^{s}(\mathbf{p})$ in the standard
representation of the gamma matrices. We have
\begin{align}
\boldsymbol{\lambda}_{\{-\text{ }+\}}^{s}(\mathbf{p})  &
=S\boldsymbol{\lambda}_{\{-\text{ }+\}}^{\prime s}(\mathbf{p})=\frac{1}%
{\sqrt{2}}\left(
\begin{array}
[c]{cc}%
\boldsymbol{1} & \mathbf{1}\\
\mathbf{1} & \boldsymbol{-1}%
\end{array}
\right)  \left(
\begin{array}
[c]{c}%
\mathbf{\sigma}_{2}[\mathbf{\phi}_{L}^{+}(\mathbf{p})]^{\ast}\\
\mathbf{\phi}_{L}^{\mathbf{+}}(\mathbf{p})
\end{array}
\right) \nonumber\\
&  =\frac{1}{\sqrt{2}}\left(
\begin{array}
[c]{c}%
\mathbf{\sigma}_{2}[\mathbf{\phi}_{L}^{+}(\mathbf{p})]^{\ast}+\mathbf{\phi
}_{L}^{\mathbf{+}}(\mathbf{p})\\
\mathbf{\sigma}_{2}[\mathbf{\phi}_{L}^{+}(\mathbf{p})]^{\ast}-\mathbf{\phi
}_{L}^{\mathbf{+}}(\mathbf{p})
\end{array}
\right)  \label{heli?}%
\end{align}
and then%
\begin{equation}
\mathbf{\Sigma}^{\prime}\cdot\frac{\mathbf{p}}{\left\vert \mathbf{p}%
\right\vert }\frac{1}{\sqrt{2}}\left(
\begin{array}
[c]{c}%
\mathbf{\sigma}_{2}[\mathbf{\phi}_{L}^{+}(\mathbf{p})]^{\ast}+\mathbf{\phi
}_{L}^{\mathbf{+}}(\mathbf{p})\\
\mathbf{\sigma}_{2}[\mathbf{\phi}_{L}^{+}(\mathbf{p})]^{\ast}-\mathbf{\phi
}_{L}^{\mathbf{+}}(\mathbf{p})
\end{array}
\right)  =\frac{1}{\sqrt{2}}\left(
\begin{array}
[c]{c}%
-\mathbf{\sigma}_{2}[\mathbf{\phi}_{L}^{+}(\mathbf{p})]^{\ast}+\mathbf{\phi
}_{L}^{\mathbf{+}}(\mathbf{p})\\
-\mathbf{\sigma}_{2}[\mathbf{\phi}_{L}^{+}(\mathbf{p})]^{\ast}-\mathbf{\phi
}_{L}^{\mathbf{+}}(\mathbf{p})
\end{array}
\right)  . \label{heli??}%
\end{equation}
\emph{Eq.(\ref{heli?})} and \emph{Eq.(\ref{heli??}) show that the labels
}$\{-$ $+\}$ \emph{(}and also $\{+$ $-\}$\emph{)} as defining the helicities
of the upper and down $\mathbb{C}^{2}$-valued components of a
$\boldsymbol{\lambda}$ type spinor field in the standard representation of the
gamma matrices have no meaning at all.\smallskip
\end{remark}

Also, one can make the identifications \footnote{See Eq.(B.6) and Eq.(B.7) in
\cite{ag2005}.}
\begin{align}
\boldsymbol{\rho}_{\{+-\}}^{s}(\mathbf{p})  &  =+i\boldsymbol{\lambda
}_{\{+-\}}^{a}(\mathbf{p}),\text{ \ \ \ \ }\boldsymbol{\rho}_{\{-+\}}%
^{s}(\mathbf{p})=-i\boldsymbol{\lambda}_{\{-+\}}^{a}(\mathbf{p}),\label{8}\\
\boldsymbol{\rho}_{\{+-\}}^{a}(\mathbf{p})  &  =-i\boldsymbol{\lambda
}_{\{+-\}}^{s}(\mathbf{p}),\text{ \ \ \ \ }\boldsymbol{\rho}_{\{-+\}}%
^{a}(\mathbf{p})=+i\boldsymbol{\lambda}_{\{-+\}}^{s}(\mathbf{p}).\nonumber
\end{align}

Moreover, we recall that the elko spinor fields are not eigenspinors of the
parity operator and indeed (see Eq.(4.14) and Eq.(4.15) in \cite{ag2005}),%

\begin{align}
\mathbf{P}\boldsymbol{\lambda}_{\{-+\}}^{s}(\mathbf{p})  &
=+i\boldsymbol{\lambda}_{\{+-\}}^{a}(\mathbf{p})=\boldsymbol{\rho}%
_{\{+-\}}^{s}(\mathbf{p}),\text{ \ }\mathbf{P}\boldsymbol{\lambda}%
_{\{+-\}}^{s}(\mathbf{p})=-i\boldsymbol{\lambda}_{\{-+\}}^{a}(\mathbf{p}%
)=\boldsymbol{\rho}_{\{-+\}}^{s}(\mathbf{p}),\nonumber\\
\mathbf{P}\boldsymbol{\lambda}_{\{-+\}}^{a}(\mathbf{p})  &
=-i\boldsymbol{\lambda}_{\{+-\}}^{s}(\mathbf{p})=\boldsymbol{\rho}%
_{\{+-\}}^{a}(\mathbf{p}),\text{\ \ }\mathbf{P}\boldsymbol{\lambda}%
_{\{+-\}}^{a}(\mathbf{p})=+i\boldsymbol{\lambda}_{\{-+\}}^{s}(\mathbf{p}%
)=\boldsymbol{\rho}_{\{-+\}}^{a}(\mathbf{p}). \label{9}%
\end{align}

Then if $\boldsymbol{\lambda}^{s,a}(x):=\boldsymbol{\lambda}^{s,a}%
(\mathbf{p})\exp(\epsilon^{s,a}ip_{\mu}x^{\mu})$, with $\epsilon^{s}=-1$ and
$\epsilon^{a}=+1$ we have due to their construction that the elko spinor
fields must satisfy the following \emph{csfopde}:%

\begin{align}
i\boldsymbol{\gamma}^{\mu}\partial_{\mu}\boldsymbol{\lambda}_{\{-+\}}%
^{s}+m\boldsymbol{\rho}_{\{+-\}}^{a}  &  =0,\text{ \ \ }i\boldsymbol{\gamma
}^{\mu}\partial_{\mu}\boldsymbol{\rho}_{\{-+\}}^{a}+m\boldsymbol{\lambda
}_{\{+-\}}^{s}=0,\nonumber\\
i\boldsymbol{\gamma}^{\mu}\partial_{\mu}\boldsymbol{\lambda}_{\{-+\}}%
^{a}-m\boldsymbol{\rho}_{\{+-\}}^{s}  &  =0,\text{ \ \ }i\boldsymbol{\gamma
}^{\mu}\partial_{\mu}\boldsymbol{\rho}_{\{-+\}}^{s}-m\boldsymbol{\lambda
}_{\{+-\}}^{a}=0,\nonumber\\
i\boldsymbol{\gamma}^{\mu}\partial_{\mu}\boldsymbol{\lambda}_{\{+-\}}%
^{s}-m\boldsymbol{\rho}_{\{-+\}}^{a}  &  =0,\text{ \ \ }i\boldsymbol{\gamma
}^{\mu}\partial_{\mu}\boldsymbol{\rho}_{\{+-\}}^{a}-m\boldsymbol{\lambda
}_{\{-+\}}^{s}=0,\nonumber\\
i\boldsymbol{\gamma}^{\mu}\partial_{\mu}\boldsymbol{\lambda}_{\{+-\}}%
^{a}+m\boldsymbol{\rho}_{\{-+\}}^{s}  &  =0,\text{ \ \ }i\boldsymbol{\gamma
}^{\mu}\partial_{\mu}\boldsymbol{\rho}_{\{+-\}}^{s}+m\boldsymbol{\lambda
}_{\{-+\}}^{a}=0. \label{10}%
\end{align}

If $\lambda_{\{+-\}}^{s,a},\lambda_{\{-+\}}^{s,a},$ $\rho_{\{+-\}}^{s,a}%
,\rho_{\{-+\}}^{s,a}\in\sec\mathcal{C\ell}^{0}(M,\eta)$ are the
representatives of the covariant spinors $\boldsymbol{\lambda}_{\{+-\}}^{s,a}%
$, $\boldsymbol{\lambda}_{\{-+\}}^{s,a}$, $\boldsymbol{\rho}_{\{+-\}}^{s,a}$,
$\boldsymbol{\rho}_{\{-+\}}^{s,a}:M\rightarrow\mathbb{C}^{4}$ then they
satisfy the \emph{csfopde}:%

\begin{align}
\boldsymbol{\partial}\lambda_{\{-+\}}^{s}\gamma_{21}+m\rho_{\{+-\}}^{a}%
\gamma_{0}  &  =0,\text{ \ \ }\boldsymbol{\partial}\rho_{\{-+\}}^{a}%
\gamma_{21}+m\lambda_{\{+-\}}^{s}\gamma_{0}=0,\nonumber\\
\boldsymbol{\partial}\lambda_{\{-+\}}^{a}\gamma_{21}-m\rho_{\{+-\}}^{s}%
\gamma_{0}  &  =0,\text{ \ \ }\boldsymbol{\partial}\rho_{\{-+\}}^{s}%
\gamma_{21}-m\lambda_{\{+-\}}^{a}\gamma_{0}=0,\nonumber\\
\boldsymbol{\partial}\lambda_{\{+-\}}^{s}\gamma_{21}-m\rho_{\{-+\}}^{a}%
\gamma_{0}  &  =0,\text{ \ \ }\boldsymbol{\partial}\rho_{\{+-\}}^{a}%
\gamma_{21}-m\lambda_{\{-+\}}^{s}\gamma_{0}=0,\nonumber\\
\boldsymbol{\partial}\lambda_{\{+-\}}^{a}\gamma_{21}+m\rho_{\{-+\}}^{s}%
\gamma_{0}  &  =0,\text{ \ \ }\boldsymbol{\partial}\rho_{\{+-\}}^{s}%
\gamma_{21}+m\lambda_{\{-+\}}^{a}\gamma_{0}=0. \label{11}%
\end{align}

\begin{remark}
From \emph{Eq.(\ref{11})} it follows trivially that the operator spinor fields
$\lambda_{\{+-\}}^{s,a},$\newline$\lambda_{\{-+\}}^{s,a}$ $\rho_{\{+-\}}%
^{s,a},\rho_{\{-+\}}^{s,a}\in\sec\mathcal{C\ell}^{0}(M,\eta)$ satisfy
Klein-Gordon equations. However, e.g., the Klein-Gordon equations
\begin{equation}
\boldsymbol{\square}\lambda_{\{-+\}}^{s}+m^{2}\lambda_{\{-+\}}^{s}=0,\text{
\ \ }\boldsymbol{\square}\rho_{\{+-\}}^{a}+m^{2}\rho_{\{+-\}}^{a}=0,
\label{14}%
\end{equation}
possess \emph{(}as it is trivial to verify\emph{)} solutions that are not
solutions of the csfopde satisfied $\lambda_{\{-+\}}^{s}$and $\rho
_{\{+-\}}^{a}$. An immediate consequence of this observation is that
\ attribution of mass dimension $1$ to elko spinor fields seems equivocated.
elko spinor fields as Dirac spinor fields have mass dimension $3/2$, and the
equation of motion for the elkos can be obtained from a Lagrangian
\emph{(}where the mass dimension of the fields are obvious\emph{)} as we
recall next.
\end{remark}

\section{Lagrangian for the \emph{csfopde} for the elko Spinor Fields}

A (multiform) Lagrangian that gives the Eqs.(\ref{11}) for the operator elko
spinor fields\ $\lambda_{\{-+\}}^{s},\lambda_{\{-+\}}^{a},$ $\rho_{\{+-\}}%
^{a},\rho_{\{-+\}}^{s}\in\sec\mathcal{C\ell}^{0}(M,\eta)$ having \emph{mass
dimension} $3/2$ is:%

\begin{equation}
\mathcal{L}=\frac{1}{2}\left\{
\begin{array}
[c]{c}%
(\boldsymbol{\partial}\lambda_{\{+-\}}^{s}\mathbf{i}\gamma_{3})\cdot
\lambda_{\{+-\}}^{s}+(\boldsymbol{\partial}\lambda_{\{-+\}}^{a}\mathbf{i}%
\gamma_{3})\cdot\lambda_{\{-+\}}^{a}+(\boldsymbol{\partial}\rho_{\{+-\}}%
^{a}\mathbf{i}\gamma_{3})\cdot\rho_{\{+-\}}^{a}\\
+(\boldsymbol{\partial}\rho_{\{-+\}}^{s}\gamma\mathbf{i}\gamma_{3})\cdot
\rho_{\{-+\}}^{s}-2m\lambda_{\{+-\}}^{s}\cdot\rho_{\{-+\}}^{a}+2m\lambda
_{\{-+\}}^{a}\cdot\rho_{\{+-\}}^{s}%
\end{array}
\right\}  \label{15}%
\end{equation}

The Euler-Lagrange equation \ obtained, e.g.,\ from the variation of the field
$\lambda_{\{+-\}}^{S}$ is\footnote{See details and the definition of the
multiform derivatives $\mathbb{\partial}_{\lambda_{\{-+\}}^{s}}$ and
$\mathbb{\partial}_{\boldsymbol{\partial}\lambda_{\{-+\}}^{s}}$in Chapters 2
and 7 of \cite{rodcap2007}.}:%
\begin{equation}
\mathbb{\partial}_{\lambda_{\{+-\}}^{S}}\mathcal{L}-\boldsymbol{\partial
}\left(  \mathbb{\partial}_{\boldsymbol{\partial}\lambda_{\{+-\}}^{s}%
}\mathcal{L}\right)  =0. \label{16}%
\end{equation}

We have immediately\footnote{In the second line of Eq.(\ref{17}) we used the
identity $(KL)\cdot M=K\cdot(M\tilde{L})$ for all $K,L,M\in\sec\mathcal{C}%
\ell(M,\eta).$}
\begin{align}
\mathbb{\partial}_{\lambda_{\{+-\}}^{s}}\mathcal{L}  &  =\frac{1}%
{2}\boldsymbol{\partial}\lambda_{\{+-\}}^{s}\mathbf{i}\gamma_{3}%
-m\rho_{\{-+\}}^{a},\nonumber\\
\mathbb{\partial}_{\boldsymbol{\partial}\lambda_{\{-+\}}^{s}}\mathcal{L}  &
=\mathcal{-}\frac{1}{2}\mathbb{\partial}_{\boldsymbol{\partial}\lambda
_{\{+-\}}^{s}}\left(  \boldsymbol{\partial}\lambda_{\{-+\}}^{s})\cdot
\lambda_{\{+-\}}^{s}\mathbf{i}\gamma_{3}\right)  =-\frac{1}{2}\lambda
_{\{+-\}}^{s}\mathbf{i}\gamma_{3},\nonumber\\
\mathcal{-}\boldsymbol{\partial}\left(  \mathbb{\partial}%
_{\boldsymbol{\partial}\lambda_{\{+-\}}^{s}}\mathcal{L}\right)   &  =+\frac
{1}{2}\boldsymbol{\partial}\lambda_{\{+-\}}^{s}\mathbf{i}\gamma_{3}.
\label{17}%
\end{align}
Recalling that $\mathbf{i}\gamma_{3}=-\gamma_{0}\gamma_{1}\gamma_{2}$ the
resulting Euler-Lagrange equation is
\[
\boldsymbol{\partial}\lambda_{\{+-\}}^{s}\gamma_{21}-m\rho_{\{-+\}}^{a}%
\gamma_{0}=0.
\]

\begin{remark}
With this result and the one in\emph{ \cite{caro2012}} we must say that the
main claims concerning the attributes of elko spinor fields appearing in
recent literature seems to us equivocated and the question arises: which kind
of particles are described by these fields and to which gauge field do they
couple? This question is answered in the next section.
\end{remark}

\section{Coupling of the elko Spinor Fields a $su(2)\simeq spin_{3,0}$ valued
Potential $\mathcal{A}$}

We start by introducing Clifford valued differential multiforms fields, i.e.,
the objects%
\begin{align}
\mathcal{K}  &  =\lambda_{\{-+\}}^{s}\otimes1-\rho_{\{+-\}}^{a}\otimes
\mathfrak{i}\tau_{2}\in\sec\mathcal{C\ell}^{0}(M,\eta)\otimes\mathbb{R}%
_{1,3}^{0}\subset\sec\mathcal{C\ell}(M,\eta)\otimes\mathbb{R}_{1,3}%
^{0}\nonumber\\
\mathcal{M}  &  =\lambda_{\{-+\}}^{s}\otimes1-\rho_{\{+-\}}^{a}\otimes
\mathfrak{i}\tau_{2}\in\sec\mathcal{C\ell}^{0\subset}(M,\eta)\otimes
\mathbb{R}_{1,3}^{0}\subset\sec\mathcal{C\ell}(M,\eta)\otimes\mathbb{R}%
_{1,3}^{0} \label{ceA0}%
\end{align}
where $\tau_{\mathbf{1},}$ $\tau_{\mathbf{2},},\tau_{\mathbf{3}}$ are the
generators of the Pauli algebra $\mathbb{R}_{3,0}\simeq\mathbb{R}_{1,3}^{0}$
and $\mathfrak{i:}=\tau_{1}\tau_{\mathbf{2}}\tau_{\mathbf{3}}$. So, we have
$\tau_{\mathbf{i}}:=\Gamma_{\mathbf{i}}\Gamma_{0}$ where the $\Gamma
_{\mathbf{\mu}}$ are the generators of $\mathbb{R}_{1,3}$, i.e.,
$\Gamma_{\mathbf{\mu}}\Gamma_{\mathbf{\nu}}+\Gamma_{\mathbf{\nu}}%
\Gamma_{\mathbf{\mu}}=2\eta_{\mathbf{\mu\nu}}$. Also, $\mathfrak{i:}=\tau
_{1}\tau_{\mathbf{2}}\tau_{\mathbf{3}}=\Gamma_{\mathbf{0}}\Gamma_{\mathbf{1}%
}\Gamma_{\mathbf{2}}\Gamma_{\mathbf{3}}=:\Gamma_{\mathbf{5}}$.

We define the reverse a general Clifford valued differential multiforms field%
\begin{equation}
\mathcal{N=N}^{\mathbf{0}}\otimes1+\mathcal{N}^{\mathbf{k}}\otimes
\tau_{\mathbf{k},}+\frac{1}{2}\mathcal{N}^{\mathbf{k}}\otimes\tau_{\mathbf{i}%
}\tau_{\mathbf{j}}+\frac{1}{3!}\mathcal{N}^{\mathbf{ikj}}\tau_{\mathbf{i}}%
\tau_{\mathbf{k}}\tau_{\mathbf{j}}\in\sec\mathcal{C\ell}(M,\eta)\otimes
\mathbb{R}_{1,3}^{0}, \label{cea12}%
\end{equation}
where $\mathcal{N}^{\mathbf{0}},\mathcal{N}^{\mathbf{k}},\mathcal{N}%
^{\mathbf{k}},\mathcal{N}^{\mathbf{ikj}}\in\sec\mathcal{C\ell}(M,\eta)$ by%
\begin{equation}
\widetilde{\mathcal{N}}=\widetilde{\mathcal{N}}^{\mathbf{0}}\otimes
1+\widetilde{\mathcal{N}}^{\mathbf{k}}\otimes\tau_{\mathbf{k},}+\frac{1}%
{2}\widetilde{\mathcal{N}}^{\mathbf{ij}}\otimes\tau_{\mathbf{j}}%
\tau_{\mathbf{i}}+\frac{1}{3!}\widetilde{\mathcal{N}}^{\mathbf{ijk}}%
\tau_{\mathbf{k}}\tau_{\mathbf{j}}\tau_{\mathbf{i}} \label{cea13}%
\end{equation}

Since, as well known the $\tau_{\mathbf{1},}$ $\tau_{\mathbf{2},}%
,\tau_{\mathbf{3}}$ have a matrix representation in $\mathbb{C(}2)$, namely
$\boldsymbol{\tau}_{\mathbf{1},}$ $\boldsymbol{\tau}_{\mathbf{2}%
,},\boldsymbol{\tau}_{\mathbf{3}}$, a set of Pauli matrices, we have the
correspondences
\begin{equation}
\mathcal{K\leftrightarrow}\left(
\begin{array}
[c]{cc}%
\lambda_{\{-+\}}^{s} & -\rho_{\{+-\}}^{a}\\
\rho_{\{+-\}}^{a} & \lambda_{\{-+\}}^{s}%
\end{array}
\right)  ,\text{ \ \ }\mathcal{M\leftrightarrow}\left(
\begin{array}
[c]{cc}%
\lambda_{\{+-\}}^{s} & -\rho_{\{-+\}}^{a}\\
\rho_{\{-+\}}^{a} & \lambda_{\{+-\}}^{s}%
\end{array}
\right)  \label{CEA00}%
\end{equation}

We observe moreover that
\begin{equation}
\mathbf{K}=\mathcal{K}\frac{1}{2}(1+\tau_{\mathbf{3}})\mathcal{\leftrightarrow
}\left(
\begin{array}
[c]{cc}%
\lambda_{\{-+\}}^{s} & 0\\
\rho_{\{+-\}}^{a} & 0
\end{array}
\right)  ,\text{ }\mathbf{M}=\frac{1}{2}\mathcal{K}\frac{1}{2}(1+\tau
_{\mathbf{3}})\mathcal{\leftrightarrow}\left(
\begin{array}
[c]{cc}%
\lambda_{\{+-\}}^{s} & 0\\
\rho_{\{-+\}}^{a} & 0
\end{array}
\right)  \label{ceA1}%
\end{equation}

Then, from Eqs.(\ref{11}) we can show that the $\mathcal{K}$ and $\mathcal{M}$
fields satisfy the following linear partial differential equations%
\begin{align}
\boldsymbol{\partial}\mathcal{K}\gamma_{21}-m\mathcal{K}\mathfrak{i}%
\tau_{\mathbf{2}}\gamma_{0}  &  =0,\label{CEA000a}\\
\boldsymbol{\partial}\mathcal{M}\gamma_{21}+m\mathcal{M}\mathfrak{i}%
\tau_{\mathbf{2}}\gamma_{0}  &  =0. \label{CEA000b}%
\end{align}

Indeed, $\mathcal{K}\mathfrak{i}\tau_{\mathbf{2}}=\mathfrak{i}\tau
_{\mathbf{2}}\mathcal{K}=\lambda_{\{-+\}}^{s}\otimes\mathfrak{i}\tau_{2}%
-\rho_{\{+-\}}^{a}\otimes1$, $\mathcal{M}\mathfrak{i}\tau_{\mathbf{2}%
}=\mathcal{M}\mathfrak{i}\tau_{\mathbf{2}}=\lambda_{\{-+\}}^{s}\otimes
\mathfrak{i}\tau_{2}-\rho_{\{-+\}}^{a}\otimes1$ and we have the
correspondences:%
\begin{align}
\mathcal{K}  &  \mathcal{\leftrightarrow}\left(
\begin{array}
[c]{cc}%
\lambda_{\{-+\}}^{s} & -\rho_{\{+-\}}^{a}\\
\rho_{\{+-\}}^{a} & \lambda_{\{-+\}}^{s}%
\end{array}
\right)  ,\text{ \ \ }\mathcal{M\leftrightarrow}\left(
\begin{array}
[c]{cc}%
\lambda_{\{+-\}}^{s} & -\rho_{\{-+\}}^{a}\\
\rho_{\{-+\}}^{a} & \lambda_{\{+-\}}^{s}%
\end{array}
\right)  ,\nonumber\\
\mathcal{K}\mathfrak{i}\tau_{\mathbf{2}}  &  =\mathfrak{i}\tau_{\mathbf{2}%
}\mathcal{K\leftrightarrow}\left(
\begin{array}
[c]{cc}%
\rho_{\{+-\}}^{a} & \lambda_{\{-+\}}^{s}\\
-\lambda_{\{-+\}}^{s} & \rho_{\{+-\}}^{a}%
\end{array}
\right)  ,\text{ \ \ }\mathcal{M}\mathfrak{i}\tau_{\mathbf{2}}=\mathfrak{i}%
\tau_{\mathbf{2}}\mathcal{M\leftrightarrow}\left(
\begin{array}
[c]{cc}%
\rho_{\{-+\}}^{a} & \lambda_{\{+-\}}^{s}\\
-\lambda_{\{+-\}}^{s} & \rho_{\{-+\}}^{a}%
\end{array}
\right)  \label{CeA04}%
\end{align}

Then, from Eqs.(\ref{CEA000a})\ and (\ref{CEA000b})\ we see that $\mathbf{K}$
and $\mathbf{M}$ satisfy the following linear partial differential equations
\begin{align}
\boldsymbol{\partial}\mathbf{K}\gamma_{21}+\mathfrak{i}m\tau_{\mathbf{2}%
}\mathbf{K}\gamma_{0}  &  =0,\label{ceA1a}\\
\boldsymbol{\partial}\mathbf{M}\gamma_{21}-\mathfrak{i}m\mathbf{\tau
}_{\mathbf{2}}\mathbf{M}\gamma_{0}  &  =0, \label{ceA2b}%
\end{align}
which, on taking the corresponding matrix representation gives the coupled
equations for the pairs $(\lambda_{\{-+\}}^{s},\rho_{\{+-\}}^{a})$ and
$(\lambda_{\{+-\}}^{s},\rho_{-+\}}^{a})$ appearing in Eqs.(\ref{11}).

Before proceeding we observe that the currents%
\begin{align}
\boldsymbol{J}_{\mathcal{K}}  &  =\mathcal{K}\tau_{\mathbf{1}}\gamma
_{0}\widetilde{\mathcal{K}}\in\sec%
{\textstyle\bigwedge\nolimits^{1}}
T^{\ast}M\otimes spin_{3,0}\hookrightarrow\sec\mathcal{C\ell}(M,\eta
)\otimes\mathbb{R}_{1,3}^{0},\label{cea6}\\
\boldsymbol{J}_{\mathcal{M}}  &  =\mathcal{M}\tau_{\mathbf{1}}\gamma
_{0}\widetilde{\mathcal{M}}\in\sec%
{\textstyle\bigwedge\nolimits^{1}}
T^{\ast}M\otimes spin_{3,0}\hookrightarrow\sec\mathcal{C\ell}(M,\eta
)\otimes\mathbb{R}_{1,3}^{0}, \label{cea7}%
\end{align}
are conserved, i.e.,
\begin{equation}
\boldsymbol{\partial}\lrcorner\boldsymbol{J}_{\mathcal{K}}=0\text{,
\ \ }\boldsymbol{\partial}\lrcorner\boldsymbol{J}_{\mathcal{M}}=0.
\label{cea8}%
\end{equation}

Indeed, let us show that $\boldsymbol{\partial}\lrcorner\boldsymbol{J}%
_{\mathcal{K}}=0$. We have%
\begin{equation}
\boldsymbol{\partial}\lrcorner\boldsymbol{J}_{\mathcal{K}}=\frac{1}{2}\left(
\boldsymbol{\partial}\mathcal{K}\tau_{\mathbf{1}}\gamma_{0}%
\widetilde{\mathcal{K}}\mathbf{+}\mathcal{K}\tau_{\mathbf{1}}\gamma
_{0}\widetilde{\mathcal{K}}\overleftarrow{\boldsymbol{\partial}}\right)
\label{cea9}%
\end{equation}
From Eq.(\ref{CEA000a}) we have%
\begin{equation}
\boldsymbol{\partial}\mathcal{K}=\mathfrak{i}m\mathcal{K}\tau_{\mathbf{2}%
}\gamma_{012},\text{ \ \ \ }\widetilde{\mathcal{K}}%
\overleftarrow{\boldsymbol{\partial}}=\partial_{\mu}\widetilde{\mathcal{K}%
}\gamma^{\mu}=\mathfrak{i}m\gamma_{012}\tau_{\mathbf{2}}\widetilde{\mathcal{K}%
}. \label{cea10}%
\end{equation}
Then,%
\begin{align*}
\boldsymbol{\partial}\lrcorner\boldsymbol{J}_{\mathcal{K}}  &  =\frac{1}%
{2}(\mathfrak{i}m\mathcal{K}\tau_{\mathbf{2}}\gamma_{012}\tau_{\mathbf{1}%
}\gamma_{0}\widetilde{\mathcal{K}}+\mathfrak{i}m\mathcal{K}\gamma_{12}%
\tau_{\mathbf{1}}\tau_{\mathbf{2}}\widetilde{\mathcal{K}})\\
&  =\frac{\mathfrak{i}m}{2}(\mathcal{K(}\tau_{\mathbf{2}}\tau_{\mathbf{1}%
}+\tau_{\mathbf{1}}\tau_{\mathbf{2}})\gamma_{12}\widetilde{\mathcal{K}}=0.
\end{align*}

The fields $\mathcal{K}$ and $\mathcal{M}$ are electrically neutral, but they
can couple with an $su(2)\simeq spin_{3,0}\subset\mathbb{R}_{3,0}$ valued
potential
\begin{equation}
\mathcal{A}=A^{\mathbf{i}}\otimes\mathbf{\tau}_{\mathbf{i}}\in\sec%
{\textstyle\bigwedge\nolimits^{1}}
T^{\ast}M\otimes spin_{3,0}\hookrightarrow\sec\mathcal{C\ell}(M,\eta
)\otimes\mathbb{R}_{1,3}. \label{ceA3}%
\end{equation}
Indeed, we have taking into account that $\mathfrak{i}=\Gamma_{\mathbf{5}%
},\tau_{\mathbf{i}}=\Gamma_{\mathbf{i}0}$ \ that the coupling is%
\begin{align}
\boldsymbol{\partial}\mathcal{K}\gamma_{21}-m\mathcal{K\Gamma}_{5}%
\mathcal{\Gamma}_{20}\gamma_{0}+q\mathcal{\Gamma}_{5}\mathcal{AK}  &
=0,\label{ceA4a}\\
\boldsymbol{\partial}\mathcal{M}\gamma_{21}+m\mathcal{M\Gamma}_{5}%
\mathcal{\Gamma}_{20}\gamma_{0}+q\mathcal{\Gamma}_{5}\mathcal{AM}  &  =0.
\label{ceA4b}%
\end{align}

Equations (\ref{ceA4a}) and (\ref{ceA4b}) are invariant under the following
transformation of the fields and change of the basis of the $spin_{3,0}%
\subset\mathbb{R}_{1,3}^{00}$ algebra:%
\begin{gather}
\mathcal{K}\mapsto\mathcal{K}^{\prime}=e^{\Gamma_{5}q\theta^{\mathbf{i}}%
\Gamma_{\mathbf{i}0}}\mathcal{K}\mathbf{,}\text{ \ \ }\mathcal{M}%
\mapsto\mathcal{M}^{\prime}=e^{\mathcal{\Gamma}_{5}q\theta^{\mathbf{i}%
}\mathcal{\Gamma}_{\mathbf{i}0}}\mathcal{M}\mathbf{,}\nonumber\\
\mathcal{A}\mapsto\mathcal{A}^{\prime}=e^{\mathcal{\Gamma}_{5}q\theta
^{\mathbf{i}}\mathcal{\Gamma}_{\mathbf{i}0}}\mathcal{A}\text{ }%
e^{-\mathcal{\Gamma}_{5}q\theta^{\mathbf{i}}\mathcal{\Gamma}_{\mathbf{i}0}%
},\text{ \ \ \ }\Gamma_{\mathbf{i}}\mapsto\Gamma_{\mathbf{i}}^{\prime
}=e^{\mathcal{\Gamma}_{5}q\theta^{\mathbf{i}}\mathcal{\Gamma}_{\mathbf{i}0}%
}\Gamma_{\mathbf{i}}e^{-\mathcal{\Gamma}_{5}q\theta^{\mathbf{i}}%
\mathcal{\Gamma}_{\mathbf{i}0}}. \label{ceA5}%
\end{gather}

With the above result we propose that elko spinor fields of the $\lambda$ and
$\rho$ types,are the crucial ingredients permitting the existence of the
$\mathcal{K}$ and $\mathcal{M}$ fields which do not carry electric charges but
possess \emph{magnetic}\footnote{The use of the term magnetic like charge here
comes from the analogy to the possibel coupling of Weyl fields describing
massless magnetic monoples with the electromagnetic potential $A\in\sec%
{\textstyle\bigwedge\nolimits^{1}}
T^{\ast}M$. See \cite{rod2003,rodcap2007}.} like charges that couple to an
$spin_{3,0}\subset\mathbb{R}_{1,3}^{00}$ valued potential $\mathcal{A}$.

\section{Difference Between Elko and Majorana Spinor Fields}

Here we recall that a Majorana field (also in class five in Lounesto
classification\footnote{We mention Dirac spinor fields are the real type
fermion fields and that Majorana and Elko spinor fields are the imaginary type
fermion fields according to Yang and Tiomno \cite{yatio} classification of
spinor fields according to their transformation laws under parity .} and
supposedly describing a Majorana neutrino) differently from an elko spinor
field\ is supposed in some textbooks to satisfy the Dirac equation (see, e.g.,
\cite{maggiore}), even if that equation cannot be derived from a Lagrangian
(unless, as it is well known the components of Majorana fields for each $x\in
M$ are Grassmann `numbers'). The \textquotedblleft proof\textquotedblright\ in
\cite{maggiore} for the statement that a Majorana field $\boldsymbol{\psi
}_{\mathbf{M}}^{\prime}:M\rightarrow\mathbb{C}^{4}$ satisfies the Dirac
equation is as follows. That author writes that $\phi_{r}:M\rightarrow
\mathbb{C}^{2}$ and $\phi_{l}:M\rightarrow\mathbb{C}^{2}$ belonging
respectively to the carrier spaces of the representations $D^{0,1/2}$ and
$D^{1/2,0}$ of $Sl(2,\mathbb{C)}$ satisfy
\begin{align}
\boldsymbol{\sigma}^{\mu}i\partial_{\mu}\mathbf{\phi}_{r}  &  =m\mathbf{\phi
}_{l},\label{weyleqs}\\
\text{ \ \ \ }\boldsymbol{\breve{\sigma}}^{\mu}i\partial_{\mu}\mathbf{\phi
}_{l}  &  =m\mathbf{\phi}_{r}, \label{weyleqs1}%
\end{align}
with $\boldsymbol{\sigma}^{\mu}=(\boldsymbol{1,}\boldsymbol{\sigma}^{i})$ and
$\boldsymbol{\breve{\sigma}}^{\mu}=(\boldsymbol{1,}-\boldsymbol{\sigma}^{i})$
where $\boldsymbol{\sigma}^{i}(=\boldsymbol{\sigma}_{i})$ are the Pauli
matrices. From this we can see that we can write:%
\begin{equation}
i\left(
\begin{array}
[c]{cc}%
\mathbf{0} & \boldsymbol{\breve{\sigma}}^{\mu}\\
\boldsymbol{\sigma}^{\mu} & \mathbf{0}%
\end{array}
\right)  \partial_{\mu}\left(
\begin{array}
[c]{c}%
\mathbf{\phi}_{r}\\
\mathbf{\phi}_{l}%
\end{array}
\right)  =m\left(
\begin{array}
[c]{c}%
\mathbf{\phi}_{r}\\
\mathbf{\phi}_{l}%
\end{array}
\right)  . \label{weyla}%
\end{equation}
The set of matrices $\boldsymbol{\gamma}^{\prime\mu}:=\left(
\begin{array}
[c]{cc}%
\mathbf{0} & \boldsymbol{\breve{\sigma}}^{\mu}\\
\boldsymbol{\sigma}^{\mu} & \mathbf{0}%
\end{array}
\right)  $ is ( as well known) a representation of Dirac matrices in Weyl
representation,. It follows that $\boldsymbol{\psi}^{\prime}$ satisfy the
Dirac equation, i.e.,
\begin{equation}
i\boldsymbol{\gamma}^{\prime\mu}\partial_{\mu}\boldsymbol{\psi}^{\prime
}-m\boldsymbol{\psi}^{\prime}=0. \label{diracagain}%
\end{equation}
Have saying that, \cite{maggiore} defines a Majorana field (in Weyl
representation) by \
\begin{equation}
\boldsymbol{\psi}_{\mathbf{M}}^{\prime}=\left(
\begin{array}
[c]{c}%
\mathbf{\phi}_{l}\\
\mathbf{\phi}_{r}%
\end{array}
\right)  =\left(
\begin{array}
[c]{c}%
\mathbf{\phi}_{l}\\
i\boldsymbol{\sigma}^{2}\mathbf{\phi}_{l}^{\ast}%
\end{array}
\right)  , \label{maj1}%
\end{equation}
and write
\begin{equation}
i\boldsymbol{\gamma}^{\prime\mu}\partial_{\mu}\boldsymbol{\psi}_{\mathbf{M}%
}^{\prime}-m\boldsymbol{\psi}_{\mathbf{M}}^{\prime}=0, \label{MAJ2}%
\end{equation}
concluding his \textquotedblleft proof\textquotedblright.

Now, let us investigate more deeply that \textquotedblleft
proof\textquotedblright. First recall that writing%
\begin{equation}
\mathbf{\phi}_{r}(x)=\mathbf{\phi}_{r}(\mathbf{p})e^{\mp ip_{\mu}x^{\mu}%
},\text{ \ \ }\mathbf{\phi}_{l}(x)=\mathbf{\phi}_{l}(\mathbf{p})e^{\mp
ip_{\mu}x^{\mu}}, \label{new1}%
\end{equation}
we have from Eq.(\ref{weyleqs}) and Eq.(\ref{weyleqs1}) that
\begin{align}
(p_{0}-\boldsymbol{\sigma\cdot}\mathbf{p})\mathbf{\phi}_{r}(\mathbf{p})  &
=\pm m\mathbf{\phi}_{l}(\mathbf{p}),\label{new2}\\
(p_{0}+\boldsymbol{\sigma\cdot}\mathbf{p})\mathbf{\phi}_{l}(\mathbf{p})  &
=\pm m\mathbf{\phi}_{r}(\mathbf{p}). \label{new3a}%
\end{align}
However, if $\mathbf{\phi}_{l}(\mathbf{0})$ and $\mathbf{\phi}_{r}%
(\mathbf{0})$\ are the zero momentum fields we have (with $\varkappa$ being
the boost parameter, i.e., $\sinh\varkappa/2=\sqrt{(\gamma-1)/2)}$ with
$\gamma=1/\sqrt{1-v^{2}}$ and $\mathbf{n}$ the direction of motion) by
definition:
\begin{align}
\mathbf{\phi}_{r}(\mathbf{p})  &  :=e^{\frac{1}{2}\varkappa\cdot
\boldsymbol{\sigma}}\mathbf{\phi}_{r}(\mathbf{0})=(\cosh\varkappa
/2+\boldsymbol{\sigma\cdot}\mathbf{n}\sinh\varkappa/2)=\frac{p_{0}%
+m+\boldsymbol{\sigma\cdot}\mathbf{p}}{[2m(p_{0}+m)]^{1/2}}\mathbf{\phi}%
_{r}(\mathbf{0}),\label{new4}\\
\mathbf{\phi}_{l}(\mathbf{p})  &  :=e^{-\frac{1}{2}\varkappa\cdot
\boldsymbol{\sigma}}\mathbf{\phi}_{l}(\mathbf{0})=(\cosh\varkappa
/2-\boldsymbol{\sigma\cdot}\mathbf{n}\sinh\varkappa/2)\mathbf{\phi}%
_{l}(\mathbf{0})=\frac{p_{0}+m-\boldsymbol{\sigma\cdot}\mathbf{p}}%
{[2m(p_{0}+m)]^{1/2}}\mathbf{\phi}_{l}(\mathbf{0}). \label{new4a}%
\end{align}

We can now verify that Eq.(\ref{new4}) and Eq.(\ref{new4a}) only imply
Eq.(\ref{new2}) and Eq.(\ref{new3a}) if\footnote{That $\mathbf{\phi}%
_{l}(\mathbf{0)=\pm\phi}_{r}(\mathbf{0)}$ is a necessary condition for a
spinor field $\boldsymbol{\psi}:M\rightarrow\mathbb{C}^{4}$ to satisfy Dirac
equation can bee seem, e.g., from Eq.(2.85) and Eq.(2.86) in Ryder's book
\cite{ryder}. However, Ryder misses the possible solution $\mathbf{\phi}%
_{l}(\mathbf{0)=-\phi}_{r}(\mathbf{0)}$. \ This has been pointed by Ahluwalia
\cite{ahlfp} in his review of Ryder's book.}
\begin{equation}
\mathbf{\phi}_{l}(\mathbf{0})=\pm\mathbf{\phi}_{r}(\mathbf{0})\mathbf{.}
\label{maj3}%
\end{equation}

But this condition cannot be satisfied for a Majorana field $\boldsymbol{\psi
}_{\mathbf{M}}^{\prime}:M\rightarrow\mathbb{C}^{4}$ as defined by
\cite{maggiore} where $\mathbf{\phi}_{r}(\mathbf{0})=i\boldsymbol{\sigma}%
^{2}\mathbf{\phi}_{l}^{\ast}(\mathbf{0)}$. Indeed, writing $\mathbf{\phi}%
_{l}^{t}(\mathbf{0})=(\nu,w)$ with $v,w\in\mathbb{C}$\ we see that to have
$\mathbf{\phi}_{l}(\mathbf{0})=\pm\mathbf{\phi}_{r}(\mathbf{0})$ we need
$\nu=\omega^{\ast}\ $and\ $\omega=-\nu^{\ast}$, i.e., $\nu=\omega=0$. We
conclude that a Majorana field $\boldsymbol{\psi}_{\mathbf{M}}^{\prime
}:M\rightarrow\mathbb{C}^{4}$ \emph{cannot} satisfy the Dirac equation.

\subsection{Some Majorana Fields are Dual Helicities Objects}

Before continuing we recall also that it is a well known fact (see,
e.g.,\cite{greiner})\emph{ }that the Dirac Hamiltonian \emph{commutes} with
the operator $\mathbf{\Sigma}\cdot\mathbf{\hat{p}}$ given by Eq.(\ref{heli}%
)\emph{.} Thus any $\Psi:M\rightarrow\mathbb{C}^{4}$ satisfying Dirac equation
which is an eigenspinor of the Dirac Hamiltonian may be constructed such
that\ $\mathbf{\phi}_{l}$ and $\mathbf{\phi}_{r}$ have the same helicity.
Since\ a Majorana spinor field $\boldsymbol{\psi}_{M}^{\prime}:M\rightarrow
\mathbb{C}^{4}$ \ as defined by \cite{maggiore} does not satisfy Dirac
equation we may suspect that it is not an eigenspinor of the of the operator
$\mathbf{\Sigma}\cdot\mathbf{\hat{p}}$. And indeed this is the case, for we
now show $\mathbf{\phi}_{l}(\mathbf{0})$ and $\ \mathbf{\phi}_{r}(\mathbf{0})$
in a Majorana field\ $\boldsymbol{\psi}_{\mathbf{M}}^{\prime}:M\rightarrow
\mathbb{C}^{4}$ are not equal. Taking the momentum (without loss of
generality) in the direction of the $z$-axis (of an inertial frame)\emph{ }and
$\mathbf{\phi}_{l}^{t}(\mathbf{0})=(1,0)$ we have
\begin{equation}
\boldsymbol{\sigma}\cdot\mathbf{\hat{p}\phi}_{l}(\mathbf{0})=-\mathbf{\phi
}_{l}(\mathbf{0})\text{, \ \ }\boldsymbol{\sigma}\cdot\mathbf{\hat{p}%
(}i\mathbf{\boldsymbol{\sigma}^{2}\phi}_{l}(\mathbf{0}))=-i\boldsymbol{\sigma
}^{2}\mathbf{\phi}_{l}(\mathbf{0})\text{,} \label{maj4}%
\end{equation}
and as the elko spinor fields they are also dual helicities objects.

\begin{remark}
Keep\ also in mind that as well known even if a Majorana field is described by
a field \emph{\cite{marsud}} $\boldsymbol{\varphi}:M\rightarrow\mathbb{C}^{2}$
carrying the $D^{1/2,0}$ \emph{(}or $D^{0,1/2}$\emph{) }representation of
$Sl(2,\mathbb{C)}$ the value of the helicity obviously depends on the inertial
reference frame where the measurement is done \emph{\cite{blp,pa} }because the
helicity is invariant only under those Lorentz transformations which did not
alter the direction of $\mathbf{p}$ along which the angular momentum component
is taken.
\end{remark}

\subsection{The Majorana Currents $\boldsymbol{J}_{\mathbf{M}}$ and
$\boldsymbol{J}_{\mathbf{M}}^{5}$}

We observe moreover that if a Majorana field $\boldsymbol{\psi}_{\mathbf{M}%
}^{\prime}:M\rightarrow\mathbb{C}^{4}$\emph{ }should satisfy the Dirac
equation then Eq.(\ref{MAJ2}) should translate in the Clifford bundle
formalism as%
\begin{equation}
\boldsymbol{\partial}\psi_{\mathbf{M}}\gamma_{21}-m\psi_{\mathbf{M}}\gamma
_{0}=0, \label{maj5}%
\end{equation}
where $\psi_{\mathbf{M}}\in\sec\mathcal{C\ell}^{0}\mathcal{(}M,\eta)$. Then,
current $\boldsymbol{J}_{\mathbf{M}}=\psi_{\mathbf{M}}\gamma_{0}\tilde{\psi
}_{\mathbf{M}}\in\sec%
{\textstyle\bigwedge\nolimits^{1}}
T^{\ast}M\hookrightarrow\sec\mathcal{C\ell(}M,\eta)$ is conserved as it is
trivial to verify. Moreover, it is \emph{lightlike }(since for a class five
spinor field $\tilde{\psi}_{\mathbf{M}}\psi_{\mathbf{M}}=0$ and thus
$\boldsymbol{J}_{\mathbf{M}}\cdot\boldsymbol{J}_{\mathbf{M}}=\psi_{\mathbf{M}%
}\gamma_{0}(\tilde{\psi}_{\mathbf{M}}\psi_{\mathbf{M}})\gamma_{0}\tilde{\psi
}_{\mathbf{M}}=0$)\emph{ }but it is a non null covector field if the
components of the spinor field $\boldsymbol{\psi}_{M}^{\prime}:M\rightarrow
\mathbb{C}^{4}$ have values in $\mathbb{C}^{2}$. Indeed writing
$\boldsymbol{J}_{\mathbf{M}}=\psi_{\mathbf{M}}\gamma_{0}\tilde{\psi
}_{\mathbf{M}}=J_{\mathbf{M}}^{\mu}\gamma_{\mu}$ we see immediately that
\begin{equation}
J_{\mathbf{M}}^{0}=\boldsymbol{\bar{\psi}}_{M}^{\prime}\boldsymbol{\gamma
}^{\prime0}\boldsymbol{\psi}_{M}^{\prime}\neq0. \label{maj6a}%
\end{equation}
Also, the current
\begin{equation}
\boldsymbol{J}_{\mathbf{M}}^{5}:=\psi_{\mathbf{M}}\gamma_{3}\tilde{\psi
}_{\mathbf{M}}=(\boldsymbol{\bar{\psi}}_{M}^{\prime}\mathbb{\ }%
\boldsymbol{\gamma}^{\prime5}\boldsymbol{\gamma}^{\prime\mu}\boldsymbol{\psi
}_{M}^{\prime})\gamma_{\mu} \label{maj7}%
\end{equation}
is non null as it is easy to verify, and is also lightlike. If the Majorana
spinor field was to satisfy the Dirac equation the current $\boldsymbol{J}%
_{\mathbf{M}}^{5}$ would be also conserved, i.e., $\boldsymbol{\partial
\lrcorner}$ $\boldsymbol{J}_{\mathbf{M}}^{5}=0$. In that case we would have a
subtle question to answer: how can a massive particle have associated to it
currents $\boldsymbol{J}_{\mathbf{M}}$ and $\boldsymbol{J}_{\mathbf{M}}^{5}$
that are ligthlike? What is the meaning of these currents?

\begin{remark}
Of course, the answer to the above question from the point of view of a first
quantized theory is that a Majorana field cannot carry any electric or
magnetic charge, i.e., the physical currents $e_{\mathbf{M}}\boldsymbol{J}%
_{\mathbf{M}}$ and $q_{\mathbf{M}}\boldsymbol{J}_{\mathbf{M}}^{5}$are null
because $e_{\mathbf{M}}=q_{\mathbf{M}}=0$.
\end{remark}

\subsection{Making of Majorana Fields that Satisfy the Dirac Equation}

Is it possible to construct a Majorana spinor field that satisfies the Dirac equation?

There are two possibilities of answering yes for the above question.\smallskip

\textbf{First possibility: }As, e.g., in \cite{leite} and \cite{cahill} we
consider ab initio a Majorana field as a quantum field and which is \emph{not}
a dual helicity object. Indeed, define a Majorana \emph{quantum field} as
$\boldsymbol{\psi}_{M}^{\prime}$ as an operator valued field satisfying
Majorana condition $\boldsymbol{\psi}_{M}^{\prime c}:=-\boldsymbol{\gamma
}^{\prime2}\boldsymbol{\bar{\psi}}_{M}^{\prime\star}=\boldsymbol{\psi}%
_{M}^{\prime}$. This condition can be satisfied if we define\footnote{Here
$\boldsymbol{\psi}_{M}^{\prime\star}:=\int\frac{d^{3}\mathbf{p}}{(2\pi)^{3/2}%
}\sum_{s}(u^{\ast}(\mathbf{p,}s)a^{\dagger}(\mathbf{p,}s)e^{-ip_{\mu}x^{\mu}%
}+v^{\ast}(\mathbf{p,}s)a(\mathbf{p,}s)e^{ip_{\mu}x^{\mu}})$}%
\begin{equation}
\boldsymbol{\psi}_{M}^{\prime}(x):=\int\frac{d^{3}\mathbf{p}}{(2\pi)^{3/2}%
}\sum_{s}(u(\mathbf{p,}s)a(\mathbf{p,}s)e^{ip_{\mu}x^{\mu}}+v(\mathbf{p,}%
s)a^{\dagger}(\mathbf{p,}s)e^{-ip_{\mu}x^{\mu}}), \label{mcl1}%
\end{equation}
with%
\begin{equation}
\{a(\mathbf{p,}s),a^{\dagger}(\mathbf{p}^{\prime}\mathbf{,}s^{\prime
})\}=\delta_{ss^{\prime}}\delta(\mathbf{p-p}^{\prime}),\text{ \ \ }%
\{a(\mathbf{p,}s),a(\mathbf{p}^{\prime}\mathbf{,}s^{\prime})\}=0 \label{mcl2}%
\end{equation}
and where the zero momentum spinors $u(\mathbf{0,}s)$ and $v(\mathbf{0,}s)$
\ are%
\begin{align}
u(\mathbf{0,}1/2)  &  =\frac{1}{\sqrt{2}}\left(
\begin{array}
[c]{c}%
1\\
0\\
1\\
0
\end{array}
\right)  ,\text{ \ \ \ }u(\mathbf{0,-}1/2)=\frac{1}{\sqrt{2}}\left(
\begin{array}
[c]{c}%
0\\
1\\
0\\
1
\end{array}
\right)  ,\nonumber\\
v(\mathbf{0,}1/2)  &  =\frac{1}{\sqrt{2}}\left(
\begin{array}
[c]{c}%
0\\
1\\
0\\
-1
\end{array}
\right)  ,\text{ \ \ \ }v(\mathbf{0,-}1/2)=\frac{1}{\sqrt{2}}\left(
\begin{array}
[c]{c}%
-1\\
0\\
1\\
0
\end{array}
\right)  , \label{mcl3}%
\end{align}
satisfying%
\begin{equation}
\boldsymbol{\gamma}_{0}^{\prime}u(\mathbf{0,}s)=u(\mathbf{0,}s),\text{
\ \ }\boldsymbol{\gamma}_{0}^{\prime}v(\mathbf{0,}s)=-v(\mathbf{0,}s)
\label{mcl4}%
\end{equation}

Indeed, we can very by explicit calculation that
\begin{equation}
u(\mathbf{p,}s)=\frac{m+p_{\mu}\boldsymbol{\gamma}^{\prime\mu}%
\boldsymbol{\gamma}^{\prime0}}{\sqrt{2p_{0}(p_{0}+m)}}u(\mathbf{0,}s),\text{
\ \ }u(\mathbf{p,}s)=\frac{m-p_{\mu}\boldsymbol{\gamma}^{\prime\mu
}\boldsymbol{\gamma}^{\prime0}}{\sqrt{2p_{0}(p_{0}+m)}}v(\mathbf{0,}s)
\label{mcl5}%
\end{equation}
and taking into account Eq.(\ref{mcl4}) we see that
\begin{equation}
(p_{\mu}\boldsymbol{\gamma}^{\prime\mu}-m)u(\mathbf{p,}s)=0,\text{
\ \ }(p_{\mu}\boldsymbol{\gamma}^{\prime\mu}+m)v(\mathbf{p,}s)=0. \label{mcl6}%
\end{equation}
With this results we can immediately verify that the quantum Majorana field
$\boldsymbol{\psi}_{M}^{\prime}(x)$ satisfy the Dirac equation,%
\begin{equation}
(i\boldsymbol{\gamma}^{\prime\mu}\partial_{\mu}-m)\boldsymbol{\psi}%
_{M}^{\prime}(x)=0. \label{mcl7}%
\end{equation}

For that Majorana field that is not a dual helicity object we can construct in
the canonical way the causal propagator that is nothing more than the standard
Feynman propagator for the Dirac equation (see, e.g.,\cite{leite}).

\begin{remark}
In a second quantized theory the currents $\boldsymbol{J}_{\mathbf{M}}$ and
$\boldsymbol{J}_{\mathbf{M}}^{5}$ are given by the normal product of the field
operators and the in this case the current $:\boldsymbol{\bar{\psi}}%
_{M}^{\prime}\boldsymbol{\gamma}^{\prime0}\boldsymbol{\psi}_{M}^{\prime}:$ as
well known is null, but $:\boldsymbol{\bar{\psi}}_{M}^{\prime}\mathbb{\ }%
\boldsymbol{\gamma}^{\prime5}\boldsymbol{\gamma}^{\prime\mu}\boldsymbol{\psi
}_{M}^{\prime}:$ is non null. For a proof of these statements, see, e.g.,
\emph{\cite{leite}} where in particular a consistent quantum field theory for
Majorana fields satisfying the Dirac equation is presented, showing in
particular that the causal propagator for that field is the standard Feynman
propagator of Dirac theory.\smallskip
\end{remark}

\textbf{Second possibility: }In several treatises, e.g., \cite{ramond,ticiati}
even at the \textquotedblleft classical level\textquotedblright\ it is
supposed that any Fermi field must be a Grassmann valued spinor field, i.e.,
an object where\ $\phi_{L}^{t}=$ $(\nu$ $\ \omega)$ and $v,\omega
:M\rightarrow\mathcal{G}$, \ with $\mathcal{G}$ a Grassmann algebra
\cite{berezin,dewitt}, i.e., $v(x)$ and $w(x)$ are Grassmann elements of a
Grassmann algebra for all $x\in M$.

In this case it is possible to show that the Majorana field defined, e.g., in
\cite{ramond} by
\begin{equation}
\mathbf{\Psi}^{\prime\mathbf{M}}=\left(
\begin{array}
[c]{c}%
\phi_{L}\\
-\boldsymbol{\sigma}_{2}\phi_{L}^{\ast}%
\end{array}
\right)  \label{grass2}%
\end{equation}
does satisfy the Dirac equation.

To prove that statement write $\phi_{L}^{t}=(%
\begin{array}
[c]{cc}%
\nu & \omega
\end{array}
)$,where for any $x\in M$, $v(x)$ and $\omega(x)$ take values in a
\emph{Grassmann algebra. }If $\mathbf{\Psi}^{\prime\mathbf{M}}$ does satisfy
Dirac equation we must have for $\mathbf{\Psi}^{\prime\mathbf{M}}(\mathbf{p})$
at $\mathbf{p=0}$ that
\begin{equation}
m\left(
\begin{array}
[c]{c}%
-\boldsymbol{\sigma}_{2}\phi_{L}^{\ast}\\
\phi_{L}%
\end{array}
\right)  -m\left(
\begin{array}
[c]{c}%
\phi_{L}\\
-\boldsymbol{\sigma}_{2}\phi_{L}^{\ast}%
\end{array}
\right)  =0. \label{grass5}%
\end{equation}
Then we need simultaneously to satisfy the equations
\begin{equation}
\phi_{L}=-\boldsymbol{\sigma}_{2}\phi_{L}^{\ast}\text{ and }\phi
_{L}=\boldsymbol{\sigma}_{2}\phi_{L}^{\ast}, \label{grass6}%
\end{equation}
which at first sight seems to be incompatible, but are not. Indeed, from
$\phi_{L}=-\boldsymbol{\sigma}_{2}\phi_{L}^{\ast}$ we obtain
\begin{equation}
\nu=i\omega^{\ast}\text{ and }\omega=-i\nu^{\ast} \label{grass3}%
\end{equation}
and from $\phi_{L}=\boldsymbol{\sigma}_{2}\phi_{L}^{\ast}$ we obtain%
\begin{equation}
\nu=-i\omega^{\ast}\text{ and }\omega=i\nu^{\ast}. \label{grass4}%
\end{equation}

So, if we \emph{understand} the symbol $\ast$ as denoting the involution
defined by Berizin's \footnote{See pages 66 and of \cite{berezin}}
Eq.(\ref{grass3}) is consistent if we take
\begin{equation}
\nu=\nu^{\ast}\text{ and }\omega=\omega^{\ast}. \label{grassy}%
\end{equation}
But since\ for any $c\in\mathbb{C}$ and $\varphi\in\mathcal{G}$ it is
$(c\varphi)^{\ast}=c^{\ast}\varphi^{\ast}$ the equation $\nu=i\omega^{\ast}$
implies%
\begin{equation}
v^{\ast}=(i\omega^{\ast})^{\ast}=-i\omega^{\ast\ast}=-i\omega\label{grassx}%
\end{equation}
and since $\nu=\nu^{\ast}$ and $\omega=\omega^{\ast}$ Eq.(\ref{grass6})
implies $\nu=-i\omega^{\ast}$. Thus, surprisingly as it may be at first sight
Eq.(\ref{grass3}) is compatible \ with Eq.(\ref{grass4}).

\begin{claim}
We may then claim that a Majorana field whose components take values in a
Grassmann algebra satisfies the Dirac equation. This is consistent with the
fact that Dirac equation under these conditions may be derived from a
Lagrangian \emph{\cite{ramond, ticiati}}. We can also verify that for such a
Majorana field the current $J_{\mathbf{M}}=0$.
\end{claim}

\begin{remark}
In resume, from the algebraic point of view there is no difference between
elko spinor fields and Majorana spinor fields $\boldsymbol{\psi}_{\mathbf{M}%
}^{\prime}:M\rightarrow\mathbb{C}^{4}$. However have in mind that the Majorana
field defined in \emph{\cite{ramond}} \ \emph{(Eq.(\ref{grass2})
}above\emph{)} looks like an elko spinor field, but, of course, is not the
same object, since the components of an elko spinor fields are for any $x\in
M$ complex numbers but the components of $\mathbf{\Psi}^{\prime\mathbf{M}}$
in\emph{ \cite{ramond}} take values in a Grassmann algebra.

Of course, if we recall that in building a quantum field theory for elkos make
automatically the components of elko spinor fields objects taking values in a
Grassmann algebra we cannot see any reason for the building of a theory like
in \emph{\cite{ag2005}}. Instead we think that elko spinor fields are worth
objects of study because they permit the construction of the $\mathcal{K}$ and
$\mathcal{M}$ fields introduced above which may describe possible of
\emph{"}magnetic like\emph{" }particles.
\end{remark}

\section{The Causal Propagator for the $\mathcal{K}$ and $\mathcal{M}$ Fields}

We now calculate the causal propagator $\mathcal{S}_{F}(x-x^{\prime})$ for,
e.g., the $\mathcal{\check{K}\in}\sec\mathcal{C\ell}^{0}\mathcal{(}%
M,\boldsymbol{\eta})\otimes\mathbb{R}_{1,3}^{0}$ field. Recall from Remark
\ref{multi} that the $\mathcal{\check{K}}$ field must satisfy
\begin{equation}
\boldsymbol{\check{\partial}}\mathcal{\check{K}}\boldsymbol{e}_{21}%
-m\mathcal{\check{K}}\Gamma_{5}\Gamma_{20}\boldsymbol{e}_{0}+\Gamma
_{5}q\mathcal{\check{A}\check{K}}=0. \label{c1}%
\end{equation}
If $\mathcal{\check{K}}_{i}(x)$ is a solution of the homogeneous equation%
\[
\boldsymbol{\check{\partial}}\mathcal{\check{K}}_{i}\boldsymbol{e}%
_{21}-m\mathcal{\check{K}}_{i}\Gamma_{5}\Gamma_{20}\boldsymbol{e}_{0}=0,
\]
we can rewrite Eq.(\ref{c1})as an integral equation%
\begin{equation}
\mathcal{\check{K}}(x)=\mathcal{\check{K}}_{i}(x)+q\int d^{4}y\mathcal{S}%
_{F}(x,y)\mathcal{\check{A}(}y\mathcal{)\check{K}(}y\mathcal{)}\Gamma
_{5}\Gamma_{20}\Gamma_{5}. \label{c11}%
\end{equation}
Putting Eq.(\ref{c11}) in Eq.(\ref{c1}) we see that $\mathcal{S}_{F}(x,y)$
must satisfy for an arbitrary $\mathcal{\check{P}\in}\sec\mathcal{C\ell
(M},\boldsymbol{\eta})\otimes\mathbb{R}_{1,3}^{0}$
\begin{equation}
\boldsymbol{\check{\partial}}\mathcal{S}_{F}(x-y)\mathcal{\check{P}%
(}y)\boldsymbol{e}_{21}-m\mathcal{S}_{F}(x-y)\mathcal{\check{P}(}%
y)\boldsymbol{e}_{0}=\delta^{4}(x-y)\mathcal{\check{P}(}y) \label{c2}%
\end{equation}
whose solution is \cite{gdl}
\begin{equation}
\mathcal{S}_{F}(x-y)\mathcal{\check{P}(}y)=\frac{1}{(2\pi)^{4}}%
{\textstyle\int}
d^{4}p\frac{\check{p}\mathcal{\check{P}(}y)+m\mathcal{\check{P}(}%
y)\boldsymbol{e}_{0}}{\check{p}^{2}-m^{2}}e^{-ip_{\mu}(x^{\mu}-y^{\mu})}.
\label{c4}%
\end{equation}
For the causal Feynman propagator we get with $E=p_{0}=\sqrt{\mathbf{p}%
^{2}+m^{2}}$
\begin{align}
\mathcal{S}_{F}(x-y)\mathcal{\check{K}(}x)  &  =\frac{-1}{2(2\pi)^{3}}%
\theta(t-t^{\prime})%
{\textstyle\int}
d^{3}p\frac{(\check{p}\mathcal{\check{K}(}y)+m\mathcal{\check{K}%
(}y)\boldsymbol{e}_{0})\boldsymbol{e}_{21}}{E}e^{-ip_{\mu}(x^{\mu}-y^{\mu}%
)}\nonumber\\
&  +\frac{1}{2(2\pi)^{3}}\theta(t-t^{\prime})%
{\textstyle\int}
d^{3}p\frac{(\check{p}\mathcal{\check{K}(}y)-m\mathcal{\check{K}%
(}y)\boldsymbol{e}_{0})\boldsymbol{e}_{21}}{E}e^{-ip_{\mu}(x^{\mu}-y^{\mu})}.
\label{c5}%
\end{align}

For a scattering problem defining $\mathcal{\check{K}}_{s}\mathcal{=\check{K}%
}-\mathcal{\check{K}}_{i}$ with $\mathcal{\check{K}}_{i}$ an asymptotic
in-state we get when $t\rightarrow\infty$
\begin{equation}
\mathcal{\check{K}}_{s}(x)=q\int d^{4}y%
{\textstyle\int}
d^{3}p\frac{(\check{p}\mathcal{\check{A}(}y\mathcal{)\check{K}}(y\mathcal{)}%
+m\mathcal{\check{A}(}y\mathcal{)\check{K}}(y\mathcal{)}\boldsymbol{e}%
_{0})\boldsymbol{e}_{21}}{2E}e^{-ip_{\mu}(x^{\mu}-y^{\mu})} \label{c6}%
\end{equation}
This permits to define a set of final states $\mathcal{\check{K}}_{f}$ given
by%
\begin{equation}
\mathcal{\check{K}}_{f}(x)=q\int d^{4}y\frac{(\check{p}_{f}\mathcal{\check
{A}(}y\mathcal{)\check{K}}(y\mathcal{)}+m\mathcal{\check{A}(}y\mathcal{)\check
{K}}(y\mathcal{)}\boldsymbol{e}_{0})\boldsymbol{e}_{21}}{2E_{f}}e^{-ip_{\mu
}(x^{\mu}-y^{\mu})} \label{c9}%
\end{equation}
which are plane waves solutions to the free field Dirac-Hestenes equation with
momentum $\check{p}_{f}$. Equipped with the $\mathcal{\check{K}}_{i}(x)$ and
$\mathcal{\check{K}}_{f}(x)$ we can proceed to calculate the scattering matrix
elements, Feynman rules and all that (see details if necessary in \cite{gdl}).

For the covariant $\boldsymbol{\lambda}$ and $\boldsymbol{\rho}$ fields the
causal propagator is the standard Dirac propagator $S_{F}(x-x^{\prime})$.
Indeed, it can be used to solve, e.g., the \emph{csfopde}
\begin{equation}
i\gamma^{\mu}\partial_{\mu}\boldsymbol{\lambda}_{\{-+\}}^{s}\gamma
_{21}+m\boldsymbol{\rho}_{\{+-\}}^{a}=0,\text{ \ \ }i\gamma^{\mu}\partial
_{\mu}\boldsymbol{\rho}_{\{+-\}}^{a}-m\boldsymbol{\lambda}_{\{-+\}}^{s}=0
\label{c20}%
\end{equation}
once appropriate initial conditions are given. To see this it is only
necessary to rewrite the formulas in Eq.(\ref{c20}) as
\begin{align}
i\gamma^{\mu}\partial_{\mu}\boldsymbol{\lambda}_{\{-+\}}^{s}%
-m\boldsymbol{\lambda}_{\{-+\}}^{s}  &  =-m(\boldsymbol{\lambda}_{\{-+\}}%
^{s}+\boldsymbol{\rho}_{\{+-\}}^{a})=\chi,\label{c2a}\\
i\gamma^{\mu}\partial_{\mu}\boldsymbol{\rho}_{\{+-\}}^{a}-m\boldsymbol{\rho
}_{\{+-\}}^{a}  &  =m(\boldsymbol{\lambda}_{\{-+\}}^{s}+\boldsymbol{\rho
}_{\{+-\}}^{a})=\varkappa\label{c2b}%
\end{align}

Eqs.(\ref{c2a}) and (\ref{c2b}) have solutions
\begin{align}
\lambda_{\{-+\}}^{s}(x)  &  =\int d^{4}yS_{F}(x-y)\chi,\label{c3}\\
\boldsymbol{\rho}_{\{+-\}}^{a}(x)  &  =\int d^{4}yS_{F}(x-y)\varkappa
\end{align}
once we recall \ that
\begin{equation}
(i\gamma^{\mu}\partial_{\mu}-m)S_{F}(x-y)=\delta^{4}(x-y).
\end{equation}

\section{Conclusions}

In \cite{rod2003,rodcap2007} it was shown that the massless Dirac-Hestenes
equation decouples in a pair of operator Weyl spinor fields, each one carrying
opposite magnetic like charges that couple to the electromagnetic potential
$A\in\sec%
{\textstyle\bigwedge\nolimits^{1}}
T^{\ast}M$ in a non standard way\footnote{In \cite{lochak} it is proposed that
the massless Dirac equation describe ( massless) neutrinos which carry pair of
opposite magnetic charges.} Here we proposed that the fields
$\boldsymbol{\lambda}$ and $\boldsymbol{\rho}$ serves the purpose of building
the fields $\mathcal{K}\mathbf{,}\mathcal{M}\in\sec\mathcal{C\ell}%
(M,\eta)\otimes\mathbb{R}_{1,3}^{0}$. These fields are electrically neutral
but carry \emph{magnetic} like charges\ which permit them to couple to a
$spin_{3,0}$ valued potential $\mathcal{A}\in\sec%
{\textstyle\bigwedge\nolimits^{1}}
T^{\ast}M\otimes spin_{3,0}$. If the field $\mathcal{A}$ is of short range the
particles described by the $\mathcal{K}$\textbf{ }and $\mathcal{M}$ may
interact forming something analogous to dark matter, in the sense that
they\ may form a condensate of spin zero particles with zero total magnetic
like charges that do not couple with the electromagnetic field and are thus invisible.

We obtained also the causal propagators for the $\mathcal{K}$\textbf{ }and
$\mathcal{M}$ fields, which can be used to calculate scattering matrix
elements, Feynman rules, etc.\footnote{At least, we can say that now we have
all the ingredients to formulate a quantum field theory for the $\mathcal{K}$
and $\mathcal{M}$ objects if one wish to do so.}

Before closing this paper we observe yet that elko spinor fields already
appeared in the literature before the publication of \cite{ag2005}. A history
about these objects may be found in \cite{dvalery11, dvalery12}. In those
papers a Lagrangian equivalent to Eq.(\ref{15}) written for the covariant
spinor fields $\boldsymbol{\lambda}$ and $\boldsymbol{\rho}$ is given.
However, the author of those papers did not comment that since the basic
\emph{csfopde} satisfied by the elko spinor fields is by construction the ones
given in Eq.(\ref{11}) and as a consequence these fields, contrary to the
claim of \cite{ag2005}, must have mass dimension $3/2$ and not $1.$

We recalled also that as claimed in \cite{ag2005} an elko spinor field (of
class five in Lounesto classification) does \emph{not} satisfy the Dirac
equation. According to some claims in the literature (see, e.g.,
\cite{maggiore} a Majorana spinor field $\boldsymbol{\psi}_{\mathbf{M}%
}^{\prime}:M\rightarrow\mathbb{C}^{4}$ and which is a dual helicity object
(that also belongs to class five in Lounesto classification) \emph{does}
satisfy the Dirac equation. However we showed that this claim is equivocated.
At \textquotedblleft classical level\textquotedblright\ a Majorana spinor
field can satisfy Dirac equation only if its components for any $x\in M$ take
values in a Grassmann algebra.

It is important to emphasize in order to avoid misunderstandings that the
theory presented in this paper is an alternative theory to the one originally
built in \cite{ag2005} and developed in a series of interesting and
challenging papers (see references). It differs drastically from that theory.
The main differences are that the equations satisfied by our elko spinor
fields of mass dimension $3/2$ (see Eq.(\ref{11})) and their solutions are
trivially Lorentz invariant. In the theory in \cite{ag2005} the elko spinor
fields are of mass dimension $1$ and that theory breaks Lorentz invariance
(see Appendix B). Also our theory gives a prediction of a new type of particle
that is electrically and magnetically neutral but has a magnetic like charge
which can couple with an $spin_{3,0}$ valued gauge field. The other theory
(for the best of our understanding) does not fix the nature of the field that
intermediates the interaction of the particles described by their elko spinor
fields of mass dimension 1. Of course, we do not claim that our theory is
better than the other. Which one is appropriate to describe some eventual
feature of physical reality (dark matter?) only the future will tell.

\begin{acknowledgement}
We are grateful to D. V. Ahluwalia for some heated but enlightening
discussions on the subject of this paper and to D. Grumiller for a crucial
remark. Moreover authors acknowledge in advance any comment they will receive.
\end{acknowledgement}

\appendix{}

\section{A New Representation of the Parity Operator Acting on Dirac Spinor
Fields}

Let $\mathbf{\langle}\boldsymbol{\mathring{e}}_{\mu}=\frac{\partial}%
{\partial\mathtt{x}^{\mu}}\mathbf{\rangle}$ and\ $\mathbf{\langle
}\boldsymbol{e}_{\mu}=\frac{\partial}{\partial x^{\mu}}\mathbf{\rangle}$ be
two arbitrary orthonormal frames for $TM$ and let $\Sigma_{0}=\mathbf{\langle
}\Gamma^{\mu}=d\mathtt{x}^{\mu}\mathbf{\rangle}$ and $\Sigma=\mathbf{\langle
}\gamma^{\mu}=dx^{\mu}\mathbf{\rangle}$ be the respective dual frames. Of
course, $\boldsymbol{\mathring{e}}_{0}$ and $\boldsymbol{e}_{0}$ are inertial
reference frames \cite{rodcap2007} and we suppose now that $\boldsymbol{e}%
_{0}$ is moving relative to $\boldsymbol{\mathring{e}}_{0}$ with $3$-velocity
$\mathbf{v=(v}^{1},v^{2},v^{3}\mathbf{)}$, i.e.,
\begin{equation}
\boldsymbol{e}_{0}=\frac{1}{\sqrt{1-v^{2}}}\boldsymbol{\mathring{e}}_{0}-%
{\textstyle\sum\limits_{i=1}^{3}}
\frac{v^{i}}{\sqrt{1-v^{2}}}\boldsymbol{\mathring{e}}_{i} \label{par1}%
\end{equation}

Let $\Xi_{u_{0}}$ and $\Xi_{u}$ be the spinorial frames associated with
$\Sigma_{0}$ and $\Sigma$. Consider a Dirac particle at rest in the inertial
frame $\boldsymbol{\mathring{e}}_{0}$ (take as a fiducial frame). The triplet
$(\psi_{0},\Sigma_{0},\Xi_{0})$ is the representative of the wave function of
our particle in $(\Sigma_{0},\Xi_{0})$ and of course, its representative in
$(\Sigma,\Xi)$ is $(\psi,\Sigma,\Xi)$. Now,
\begin{equation}
\psi=u\psi_{0} \label{par2}%
\end{equation}
where $u$ describes in the spinor space the boost sending $\Gamma^{\mu}$ to
$\gamma^{\mu}$, i.e.,$\gamma^{\mu}=u$ $\Gamma^{\mu}u^{-1}=\Lambda_{\nu}^{\mu
}\Gamma^{\nu}$. Now, the representative of the parity operator in $(\Sigma
_{0},\Xi_{0})$ is $\mathcal{P}_{u_{0}}$ and in $(\Sigma,\Xi)$ is
$\mathcal{P}_{u};$ We have according to our dictionary (Eq.(\ref{5})) that
\begin{equation}
\mathcal{P}_{u}\psi=\gamma^{0}\psi\gamma^{0},\text{ \ \ \ }\mathcal{P}_{u_{0}%
}\psi_{0}=\Gamma^{0}\psi_{0}\Gamma^{0}, \label{par3}%
\end{equation}
or
\begin{equation}
\mathcal{P}_{u}\mathbf{\Psi}=\gamma^{0}\mathcal{R}\mathbf{\Psi},\text{
\ \ \ }\mathcal{P}_{u_{0}}\mathbf{\Psi}_{0}=\Gamma^{0}\mathcal{R}\mathbf{\Psi
}_{0}, \label{par4}%
\end{equation}
where $\mathbf{\psi}$ and $\mathbf{\psi}_{0}$ are Dirac \emph{ideal} real
spinor fields\footnote{See \cite{rodcap2007} for details.}
\begin{equation}
\mathbf{\Psi}=\psi\frac{1}{2}(1+\gamma^{0}),\text{ \ \ }\mathbf{\Psi}_{0}%
=\psi_{0}\frac{1}{2}(1+\Gamma^{0}), \label{par5}%
\end{equation}
and if the momentum of our particle is the covector field $\boldsymbol{p}%
=\mathring{p}_{\mu}\Gamma^{\mu}=p_{\mu}\gamma^{\mu}$ with $(\mathring{p}%
_{0},\mathring{p}_{1},\mathring{p}_{2},\mathring{p}_{3}):=(m,\mathbf{0})$ and
$(p_{0},p_{1},p\mathring{p}_{2},p_{3}):=(E,\mathbf{p})$ (and of course
$p_{\mu}=$\ $\Lambda_{\mu}^{\nu}\mathring{p}_{\mu}=\Lambda_{\mu}^{0}%
\mathring{p}_{0}$) $\mathcal{R}$ an the operator such that if $\psi
=\phi(\mathbf{p})e^{i\boldsymbol{px}}$ then
\begin{equation}
\mathcal{R}\psi=\phi(\mathbf{p})e^{-i\boldsymbol{p}_{\mu}\boldsymbol{x}^{\mu}%
}=\phi(\mathbf{p})e^{-i(\boldsymbol{p}_{0}\boldsymbol{x}^{0}-p_{i}x^{i})}.
\label{par5a}%
\end{equation}
Also $u\mathcal{R=R}u$ and clearly $\mathcal{R}\psi_{0}=\psi_{0}$. Now,
\begin{equation}
u\mathcal{P}_{u_{0}}u^{-1}u\mathbf{\Psi}_{0}=u\Gamma^{0}\mathcal{R}%
\mathbf{\Psi}_{0}=u\Gamma^{0}u^{-1}\mathcal{R}u\mathbf{\Psi}_{0}=\gamma
^{0}\mathcal{R}\mathbf{\Psi,} \label{PAR6}%
\end{equation}
from where it follows that
\begin{equation}
\mathcal{P}_{u}=u\mathcal{P}_{u_{0}}u^{-1}. \label{par7}%
\end{equation}
Now we rewrite $\mathcal{P}_{u}\mathbf{\Psi}=\gamma^{0}\mathcal{R}%
\mathbf{\Psi}$ as%
\begin{align}
\mathcal{P}_{u}\mathbf{\Psi}  &  =\frac{\mathring{p}_{0}}{m}u\Gamma
^{0}\mathcal{R}\mathbf{\Psi}_{0}=\frac{\mathring{p}_{0}}{m}u\Gamma^{0}%
u^{-1}u\mathbf{\Psi}_{0},\nonumber\\
&  =\frac{\mathring{p}_{0}}{m}\Lambda_{\mu}^{0}\gamma^{\mu}\mathbf{\Psi}%
=\frac{1}{m}p_{\mu}\gamma^{\mu}\mathbf{\Psi.} \label{par8}%
\end{align}

We conclude that the parity operator in an arbitrary orthonormal and spin
frames $(\Sigma,\Xi)$ acting on a Dirac ideal spinor field $\mathbf{\psi}$) is%
\begin{equation}
\mathcal{P}=\mathcal{P}_{u}=\frac{1}{m}p_{\mu}\gamma^{\mu}. \label{par9}%
\end{equation}

Of course, when applied to\ covariant spinor fields $\boldsymbol{\psi
}:M\rightarrow\mathbb{C}^{4}$ the operator $\mathcal{P}$ is represented by
\begin{equation}
\boldsymbol{P}=\frac{1}{m}p_{\mu}\boldsymbol{\gamma}^{\mu}. \label{par10}%
\end{equation}

A derivation of this result using covariant spinor fields (and which can be
easy generalized for arbitrary\ higher spin fields) has been obtained in
\cite{sper}.

\section{Correct Value for the Fourier Transform of $\mathcal{G}(\mathbf{p})$}

According to the theory of elko spinor fields as originally developed in
\cite{ag2005} (see also \cite{ag2005a,ahlfp,ah2010,als2010,als2011,ahl2013}
the evaluation of the anticommutator of an elko spinor field with its
canonical momentum gives
\begin{equation}
\{\Lambda(\mathbf{x,}t),\Pi(\mathbf{x},t\}=i\delta(\mathbf{x}-\mathbf{x}%
^{\prime})\mathbb{I}\text{ }+i\int\frac{d^{3}p}{(2\pi)^{3}}e^{i\mathbf{p\cdot
(x-x}^{\prime})}\mathcal{G(}\mathbf{p}), \label{als 42}%
\end{equation}
with%

\begin{equation}
\mathcal{G(}\mathbf{p}):=\gamma^{5}\gamma^{\mu}n_{\mu}(\mathbf{p}),~~~
\label{als 31}%
\end{equation}
where the spacelike $\mathbf{p}$-dependent field $n=n_{\mu}(\mathbf{p}%
)\mathbf{e}_{\mu}$ is
\begin{gather}
(n_{0}(\mathbf{p}),n_{1}(\mathbf{p}),n_{2}(\mathbf{p}),n_{3}(\mathbf{p}%
)):=(0,\mathbf{n(p)),}\nonumber\\
\mathbf{p}=(p\cos\theta,p\sin\theta\cos\varphi,p\sin\theta\sin\varphi)\\
\mathbf{n(p):=}\frac{1}{\sin\theta}\frac{\partial}{\partial\varphi}\left(
\frac{\mathbf{p}}{\left\vert \mathbf{p}\right\vert }\right)  =(-\sin
\varphi,\cos\varphi,0)\nonumber\\
=\left(  -\tau(1+\tau^{2})^{-1/2},\tau(1+\tau^{2})^{-1/2},0\right)
,~~~\tau=p_{y}/p_{x}. \label{als31a}%
\end{gather}

Putting $\mathbf{\Delta=x-x}^{\prime}$ it is\footnote{This result has been
also found by Ahluwalia and Grumiller. We find the above result without
knowing their calculations, which Ahluwalia informed us will \ be reported in
a forthcoming paper.}%

\begin{align}
\mathcal{\hat{G}}(\mathbf{\Delta})  &  =\int\frac{d^{3}p}{(2\pi)^{3}%
}e^{i\mathbf{p\cdot(x-x}^{\prime})}\mathcal{G(}\mathbf{p})\nonumber\\
&  =-\gamma^{5}\gamma^{1}P(\mathbf{\Delta})+\gamma^{5}\gamma^{2}%
Q(\mathbf{\Delta}). \label{N1}%
\end{align}

with
\begin{equation}
P(\mathbf{\Delta})=-\frac{i}{2\pi}\delta(\Delta_{z})\frac{\Delta_{y}}%
{(\Delta_{x}^{2}+\Delta_{y}^{2})^{\frac{3}{2}}},~~~Q(\mathbf{\Delta})=\frac
{i}{2\pi}\delta(\Delta_{z})\frac{\Delta_{x}}{(\Delta_{x}^{2}+\Delta_{y}%
^{2})^{\frac{3}{2}}} \label{N4a}%
\end{equation}

\begin{remark}
In \emph{\cite{cr2012}} the integral in\ \emph{Eq.(\ref{N1})} has been
evaluated for the case when $\mathbf{\Delta}$ lies in the direction of
\emph{one} of the spatial axes $\mathbf{e}_{i}=\partial/\partial x^{i}$ of an
\ arbitrary inertial reference frame $\mathbf{e}_{0}=\partial/\partial x^{0}$
\emph{(}where $(x^{0},x^{1},x^{2},x^{3})$ are coordinates in
Einstein-Lorentz-Poincar\'{e} gauge naturally adapted to $\mathbf{e}_{0}$
\cite{rodcap2007,sw}\emph{)}.We note that the evaluation of each one of the
integrals in \emph{\cite{cr2012}} is correct, but they \emph{do not} express
the values of the Fourier transform $\mathcal{\hat{G}}(\mathbf{x-x}^{\prime})$
for the particular values of $\mathbf{\Delta}$ used in the calculations of
those integrals. It is not licit to fix a priori two of the components of
$\mathbf{\Delta}$ as being null to calculate the integral $(2\pi)^{-3}%
{\textstyle\int}
d^{3}pe^{i\mathbf{p\cdot(x-x}^{\prime})}\mathcal{G(}\mathbf{p})$ for this
procedure excludes the singular behavior in the sense of distributions of the
Fourier integral. So, it is wrong the statement in \emph{\cite{caro2012}} that
elko theory as constructed originally in \emph{\cite{ag2005}}\ is
local\emph{\footnote{An errata has been sent to PRD.}}.
\end{remark}

\subsection{Plane of Nonlocality and Breakdown of Lorentz Invariance}

When $\Delta_{z}\neq0$, $\mathcal{\hat{G}}(\mathbf{x-x}^{\prime})$ is null the
anticommutator is \emph{local} and thus there exists in the elko theory\ as
constructed in \cite{ag2005,als2010} an infinity number of
\emph{\textquotedblleft}locality\ directions\emph{\textquotedblright.} On the
other hand $\mathcal{\hat{G}}(\mathbf{x-x}^{\prime})$ is\ a distribution with
support in $\Delta_{z}=0$. So$,$the directions $\mathbf{\Delta}=(\Delta
_{x},\Delta_{y},0)$\ are nonlocal in each arbitrary inertial reference frame
$\mathbf{e}_{0}$ chosen to evaluate $\mathcal{\hat{G}}(\mathbf{x-x}^{\prime}%
)$. Recall that given an inertial (coordinate) reference frame frame
$\mathbf{e}_{0}=\partial/\partial x^{0}$ in Minkowski spacetime there exists
\cite{choquet} and infinity of triples of vector fields $\{\mathbf{e}%
_{1}^{(k)}=\partial/\partial x_{(k)}^{1},\mathbf{e}_{1}^{(k)}=\partial
/\partial x_{(k)}^{2},\mathbf{e}_{1}^{(k)}=\partial/\partial x_{(k)}^{3}\}$
(with $(x^{0},x_{(k)}^{1},x_{(k)}^{2},x_{(k)}^{3})$, coordinates in
Einstein-Lorentz-Poincar\'{e} gauge \cite{rodcap2007} naturally adapted to
$\mathbf{e}_{0}$ differing by a spatial rotation) which constitutes a global
section of the frame bundle. So, the labels $x$,$y$ and $z$ directions in
inertial reference frame $\mathbf{e}_{0}$ are arbitrary (a mere convention)
and thus without \emph{any} physical significance. This means that the theory
as constructed in \cite{ag2005} breaks in each inertial reference frame
rotational invariance and since in different inertial references frames there
are different $(x,y)$ planes the theory breaks also Lorentz invariance. This
means that the theory as constructed in \cite{ag2005} breaks in each inertial
reference frame rotational invariance and since in different inertial
references frames there are different ( and moreover arbitrary) $(x,y)$ planes
the theory breaks also Lorentz invariance. This odd feature (according to our
view) of the theory of elko spinor fields as constructed originally in
\cite{ag2005} was eventually the main reason that lead us to the investigation
described in this paper. However if this odd effect will be observed in
experiments we must agree that the original version of elko theory is indeed a
science breakthrough. The one who lives will know.


\begin{thebibliography}{99}                                                                                               %


\bibitem {ag2005}Ahluwalia-Khalilova, D. V. and Grumiller, D.,\ Spin-half
Fermions with Mass Dimension One: Theory, Phenomenology, and Dark Matter,
\emph{JCAP} \textbf{07} 012 (2005)\emph{ }\texttt{[hep-th/0412080]}

\bibitem {ag2005a}Ahluwalia-Khalilova, D. V. and Grumiller, D., Dark Matter: A
Spin One Half Fermion Field with Mass Dimension One?,\textit{ Phys.Rev. D}
\textbf{72} 067701 (2005) \texttt{[hep-th/0410192]}

\bibitem {ahlfp}Ahluwalia, D. V, Review of \textit{Quantum Field Theory},
Second Edition. By Lewis H. Ryder (Cambridge University Press, United Kingdom,
1996), \emph{Found. Phys}. \textbf{28}, 527-528 (1998).

\bibitem {ah2010}Ahluwalia, D. V., and Horvath, S. P., Very Special Relativity
of Dark Matter; The elko Connection, JHEP \textbf{11}, 078 (2010).
\texttt{[arXiv:1008.0436v2 [hep-ph]]}

\bibitem {als2010}Ahluwalia, D. V., Lee, C. Y., and Scritt, D, Elko as Self
Interacting Fermionic Dark Matter with Axis of Locality, \emph{Phys. Lett. B}.
\textbf{687}, 248-252 (2010).

\bibitem {als2011}Ahluwalia, D. V., Lee, C. Y., and Scritt, D.,
Self-Interacting Matter With an Axis of Locality, \emph{Phys. Rev. D}
\textbf{83}, 065017 (2011).

\bibitem {blp}Berestetski\u{\i}, V. B., Lifishitz, E. M., and
Pitaesvski\u{\i}, L., \emph{Relativistic Quantum Theory}, \emph{Part I},
Pergamon Press, New York, 1971.

\bibitem {berezin}Berezin, F. A., \emph{The Method of Second Quantization},
Academic Press, New York, 1966.

\bibitem {cahill}Cahill, P. and Cahill, K., Learning about Spin-One-Half
Fields, \emph{Eur. J. Phys}. \textbf{27,} 29-47 (2006) and corrigenda at
\emph{Eur. J. Phys}. \textbf{28,} 145 (2006).

\bibitem {caro2012}Capelas de Oliveira, E., and Rodrigues, W. A. Jr., Comment
on \textquotedblleft Self-interacting Elko Dark Matter with an Axis of
Locality\textquotedblright\ \emph{Phys. Rev. D} \textbf{86, }128501
(2012)\texttt{.}

\bibitem {choquet}Choquet-Bruhat, Y., DeWitt-Morette, C., Dillard-Bleick, M.,
\emph{Analysis, Manifolds and Physics} (revised edition), North Holland,
Amsterdam, \ 1982.

\bibitem {dewitt}D{\small e}Witt, B., Supermanifold, Cambridge University
Press, Cambridge, 1984.

\bibitem {dvalery11}Dvoeglazov, V.V., Majorana Neutrino: Chirality and
Helicity, \textit{J. Phys. Conf. Ser.} \textbf{343}, 012033 (2012).
\texttt{[arXiv:1108.4991v2 [math-ph]]}

\bibitem {dvalery12}Dvoeglazov, V.V., How to construct Self/Anti-Self Charge
Conjugate States for Higher Spins?\ \texttt{[arXiv:1210.4401v1 [math-ph]]}

\bibitem {GR}Gradshteyn, I. S. and Ryzhik, I. M., \emph{Table of Integrals
Series and Products} (fourth edition), 1965. Academic Press, New York,

\bibitem {greiner}Greiner, W., \emph{Relativistic Quantum Mechanics},
pp.84-85, Springer-Verlag, Berlin,1994.

\bibitem {gdl}Gull, S., Doran, C, and Lasenby, A., Electron Physics II, in
Baylis, W. E. (ed.), \emph{Clifford }(\emph{Geometrical})\emph{ Algebras},
pp.129-145, Birkha\"{u}ser, Boston, 1996.

\bibitem {leite}Leite Lopes, J., \emph{Lectures on Symmetries}, Gordon and
Breach, New York, 1969.

\bibitem {lochak}Lochak, G., Wave Equation for a Magnetic Monopole, \emph{Int.
J. Theor. Phys}. \textbf{24}, 1019-1050 (1985).

\bibitem {lounesto}Lounesto, P., \emph{Clifford Algebras and Spinors},
Cambridge University Press, Cambridge, 1997.

\bibitem {maggiore}Maggiore, M., \emph{A Modern Introduction to Quantum Field
Theory}, Oxford Univ. Press, Oxford, 2005.

\bibitem {marsud}Marshak, R. E. and Sudarshan, E.C.G., \textit{Introduction to
Elementary Particle Physics}, John Wiley, New York 1961.

\bibitem {mr2004}Mosna, R. A., and Rodrigues, W. A. Jr., The Bundles of
Algebraic and Dirac-Hestenes Spinor Fields, \emph{J. Math. Phys.} \textbf{45},
2945-2966 (2004). \texttt{[arXiv:math-ph/0212033]}

\bibitem {pa}Pal, P. B., Dirac, Majorana and Weyl Fermions, \emph{Am. J.
Phys.} \textbf{79}, 485-498 (2011). \texttt{[arXiv:1006.1718v2 [hep-ph]]}

\bibitem {ramond}Ramond, P.,\emph{ Field Theory: A Modern Primer} (second
edition), Addison-Wesly Publ. Co., Reading, MA, 1989.

\bibitem {r2004}Rodrigues, W. A. Jr., Algebraic and Dirac-Hestenes Spinors and
Spinor Fields, \emph{J. Math. Phys}. \textbf{45}, 2908-2994 (2004).
\texttt{[arXiv:math-ph/0212030]}

\bibitem {roro2006}da Rocha, R. and Rodrigues, W. A. Jr., Where are elko
Spinor Fields in Lounesto Spinor Field Classification?, \emph{Mod. Phys.
Lett}. \emph{A} \textbf{21, }65-74\textbf{ }(2006) \texttt{[math-ph/0506075]}

\bibitem {rh2007}da Rocha, R. and da Silva,J. M. Hoff, From Dirac Spinor
Fields to elko, \emph{J. Math. Phys}.\textbf{48} 123517
(2007).\texttt{[arXiv:0711.1103 [math-ph]].}

\bibitem {rh2010}da Rocha, R., and da Silva,J. M. Hoff, elko, Flagpole and
Flag-Dipole Spinor Fields, and the Instanton Hopf Fibration, \emph{Adv. Appl.
Clifford Algebras} \textbf{20, }847-870 (2010). \texttt{[arXiv:0811.2717
[math-ph]]}

\bibitem {rod2003}Rodrigues, W. A. Jr., The Relation Between Maxwell, Dirac
and the Seiberg-Witten Equations, \emph{Int. J. Math. and Math. Sci.}
\textbf{2003}, 2707-2734 (2003). \texttt{[arXiv:math-ph/0212034]}

\bibitem {rodcap2007}Rodrigues, W. A. Jr., and Capelas de Oliveira, E.,
\textit{The Many Faces of Maxwell, Dirac and Einstein Equations. A Clifford
Bundle Approach}. Lecture Notes in Physics \textbf{722}, Springer, Heidelberg, 2007.

\bibitem {ryder}Ryder, L. H., \textit{Quantum Field Theory} (second edition),
Cambridge Univ. Press, Cambridge, 1996.

\bibitem {sw}Sachs, R. K., and Wu, H., \textit{General Relativity for
Mathematicians,} Springer, New York, \ 1977.

\bibitem {siro2009}da Silva, J..M. Hoff, and da Rocha, R., From Dirac Action
to elko Action, \emph{Int. J. Mod. Phys. A} \textbf{24}, \textbf{ }3227-3242
(2009). \texttt{[arXiv:0903.2815 [math-ph]]}

\bibitem {sch}Schweber, S. S., \emph{An Introduction to Relativistic Quantum
Field Theory}, Harper and Row, New York, 1964.

\bibitem {sper}Speran\c{c}a, L. D., An Identification of the Dirac Operator
with the Parity Operator \texttt{[arXiv:1304.4794 [math-ph]]}

\bibitem {ticiati}Ticciati, R., \emph{Quantum Theory for Mathematicians},
Cambridge University Press, Cambridge, 1999.

\bibitem {yatio}Yang, C. N., and Tiomno, J., Reflection Properties of Spin
$1/2$\ \ Fields and a Universal Fermi-Type Interaction, \textit{Phys. Rev}.
\textbf{88}, 495-498 (1950).
\end{thebibliography}
\end{document}